\documentclass[usenatbib]{mn2e}

\usepackage{graphicx}
\usepackage{graphics}
\usepackage{amsmath}
\usepackage{amssymb}
\usepackage{color}
\usepackage{url}
\usepackage[draft]{hyperref}
\usepackage{pdflscape}
\usepackage{setspace}
\RequirePackage{fixltx2e}

\def\aap{A\&A}

\def\apj{ApJ}
\def\apjl{ApJL}
\def\apjs{ApJS}

\def\mnras{MNRAS}

\def\araa{ARAA}

\begin{document}

\title[Properties of the Simulated CGM]{\singlespacing Properties of the Simulated Circumgalactic Medium}
\author[Lochhaas et al.]{Cassandra Lochhaas$^{1,2}$, Greg L. Bryan$^{3,4}$, Yuan Li$^{4,5}$, Miao Li$^{4}$, Drummond Fielding$^{4,5}$ \\
$^{1}$ Department of Astronomy, The Ohio State University, 140 West 18th Avenue, Columbus, OH 43210, USA\\
$^{2}$ Space Telescope Science Institute, 3700 San Martin Drive, Baltimore, MD 21218, USA\\
$^{3}$ Department of Astronomy, Columbia University, 550 West 120th Street, New York, NY 10027, USA\\
$^{4}$ Center for Computational Astrophysics, Flatiron Institute, 162 5th Avenue, New York, NY 10010, USA\\
$^{5}$ Department of Astronomy and Theoretical Astrophysics Center, University of California at Berkeley, 501 Campbell Hall \#3411,\\
Berkeley, CA 94720, USA
}
\maketitle

\begin{abstract}
The circumgalactic medium (CGM) is closely linked to galaxy formation and evolution, but difficult to characterize observationally and typically poorly resolved in cosmological simulations. We use spherically-symmetric, idealized, high-resolution simulations of the CGM in $10^{12}M_\odot$ and $10^{11}M_\odot$ dark matter halos to characterize the gas pressure, turbulent and radial velocities, and degree of thermal and effective dynamic pressure support in the overall CGM as well as in its high- and low-temperature phases. We find that the $10^{12}M_\odot$ halo contains a CGM mostly formed of a hot gas halo in hydrostatic equilibrium out of which cold gas condenses and falls onto the central galaxy, while the $10^{11}M_\odot$ halo's CGM is not in hydrostatic equilibrium, has a wider spread of properties at a given galactocentric radius, does not have a clear separation of hot and cold phases, and is dominated by bulk motions. We also find that the degree of pressure support in the $10^{11}M_\odot$ halo is strongly dependent on the parameters of the galactic winds of the central galaxy. These results promote the idea that there is no ``average" CGM and care must be taken when setting the initial conditions for a small-box simulation of a patch of the CGM.
\end{abstract}

\begin{keywords}
galaxies: haloes
\end{keywords}

\section{Introduction}

The circumgalactic medium (CGM) is the conduit between galaxies and the intergalactic medium, through which all gas that flows into or out of a galaxy must pass \citep[see][for a recent review]{Tumlinson2017}. Observations have indicated that the CGM contains a significant amount of mass, perhaps as much or more as the mass in the galactic disk \citep{Peeples2014,Werk2014,Keeney2017}, and that it is multiphase with a variety of different temperatures and densities \citep[e.g.,][]{Gupta2012,Tumlinson2013,Bordoloi2014,Borthakur2015}. The large mass and close location of the CGM to the galaxies that host it indicate it is closely linked to the growth and evolution of galaxies, yet there are many open questions about the processes governing it.

Due to the diffuse nature of much of the CGM, it is difficult to characterize observationally. Typical surveys probe diffuse X-ray emission \citep{Anderson2010,Anderson2013} or require spectra of background quasars and obtain just a handful of lines of sight through a given galaxy's CGM \citep{Rudie2012,Tumlinson2013,Bordoloi2014,Borthakur2015,Keeney2017} and are limited in probing the full structure. Quasar absorption-line studies like these have revealed the existence of cool, warm, and hot gas in the CGM, sometimes all present along a single sightline, indicating that the CGM has a significant amount of structure. Observed velocities of gas in the CGM range from near the systemic velocity of the galaxy to hundreds of kilometers per second, suggesting the kinematic structure is just as important as the spatial structure. Theoretical and observational estimated sizes of the cool clouds range from $\sim1$ pc \citep[e.g.,][]{McCourt2012,Liang2018} to $100$ kpc \citep[e.g.,][]{Stocke2013,Werk2014}, suggesting that potentially very high spatial resolution is needed to understand the structure of the CGM. Quasar absorption line observations aim to produce a statistical sampling of the ``typical" CGM.

Similarly, analytic works that aim to determine an overall unifying picture of the physics of the CGM, and in particular its multiphase nature, predict a wide variety of cloud sizes, kinematic structure, and physical processes. Models of galactic winds interacting with the CGM can produce multiphase gas at a variety of velocities through shock-heating and radiative cooling \citep{Samui2008,FaucherGiguere2012,Lochhaas2018,McQuinn2018}. Models of cool gas condensing out of a hot medium predict formation times and lifetimes of cool clouds and pressure balance between hot and cool gas and overall pressure support of the halo \citep{McCourt2012,Faerman2017,Voit2017,Liang2018,McCourt2018,Faerman2019,Voit2019}. However, analytic models tend to be idealized and neglect certain physical processes, like the evolution of dynamic or sheer instabilities, in order to be analytically tractable.

Instead, simulations can provide additional insight into not just the statistical properties of the CGM, but also its physical structure and evolution, including many physical processes not included in analytic models. Fully cosmological simulations do not have high enough resolution to resolve the CGM, and cosmological zoom-in simulations \citep{Angles-Alcazar2014,Hopkins2014,Rahmati2016,Henden2018,Pillepich2018} typically refine the densest regions, the galaxies, while the diffuse CGM remains poorly resolved. Simulations specifically designed to study the CGM at high resolution are needed \citep{Sarkar2015,Fielding2017,Hummels2018,Peeples2019,vandeVoort2019}. \citet{Peeples2019} showed that higher resolution simulations of the CGM reveal more detailed structure, especially in the cool clouds predominantly probed by quasar absorption line studies.

Simulations of galaxy clusters and massive galaxies found that cold gas condenses out of a hot medium if the cooling time is short compared to the dynamical time of the gas \citep{McCourt2012,Sharma2012,Li2014,Choudhury2019,Wang2019}, and it is possible that similar models apply to lower-mass galaxy halos as well \citep{Voit2019}. Such instability could be a source of the cool phase of gas in the CGM, but extremely high-resolution simulations are necessary to resolve the cooling length. The CGM extends to the dark matter halo virial radius, hundreds of kiloparsecs for massive galaxies, so resolving the entirety of the CGM at the level required to fully track condensation is computationally expensive. Small-box simulations of just a small region of a galactic wind flow or the CGM \citep{Ferrara2016,Gronke2018,Liang2018} can reach the resolution needed to track cooling condensation, but they must be initialized with certain properties that represent a realistic CGM, such as a general gas pressure or turbulent and outflowing velocities. Full-CGM simulations, while not high enough resolution to resolve the coldest clouds, are necessary for providing the overall properties of the CGM that can be used for initializing small-box simulations. In addition, idealized simulations of the CGM are useful for understanding the bulk physics affecting the gas.

In this paper, we aim to quantify not just the overall mean properties of the simulated CGM, but also the spatial, temporal, and statistical fluctuations of the gas. Understanding the physical processes that produce and govern all parts of the multiphase gas, not just the majority of it, provides the specificity needed for initializing small-box CGM simulations and provides a basis from which to interpret observations that trace different phases of gas. We base our analysis on the simulations of \citet{Fielding2017}, which are high-resolution, 3D, idealized, isolated galaxy simulations of the full extent of the CGM.


In \S\ref{sec:sim}, we give a brief overview of the simulations used and how we analyze them. Section~\ref{sec:results} presents our findings of pressure and velocity distributions at a given radius (\S\ref{sec:hist}), and of how the pressure (\S\ref{sec:pres}), velocity (\S\ref{sec:vel}), turbulence (\S\ref{sec:turb}), and pressure support (\S\ref{sec:HSE}) vary with galactocentric radius. We discuss the picture of how the CGM changes with halo mass (\S\ref{sec:halomass}), the impact of our results for small-box simulations (\S\ref{sec:smallbox}), and how the results vary with the implementation of the galactic wind (\S\ref{sec:feedback}) in \S\ref{sec:discussion}, and summarize and give conclusions in \S\ref{sec:summary}.

\section{Simulations and Analysis}
\label{sec:sim}

We use the simulations of \citet{Fielding2017}, and refer the reader to that study for more details but give an overview of the simulations here. These are 3D hydrodynamic simulations using the ATHENA code that do not model the galaxy nor cosmological structure. They have static mesh refinement so that the region closest to the galaxy at the center of the domain has higher resolution than the outskirts of the CGM; for the larger- (smaller-)mass halo simulation (see below) the resolution varies from 1.4 (0.65) kpc close to the galaxy to 5.6 (2.6) kpc at the outer edge of the domain, past the virial radius of the halo. They are spherically symmetric, with the inner edge of the domain at small radius representing the (spherical) galaxy. The only gravity included is that of a NFW \citep{Navarro1997} dark matter halo.

Both the inner edge and the outer edge allow inflows and outflows --- at the outer edge, there is spherical cosmological accretion at a rate of $\dot M_\mathrm{acc}=7 M_\odot$ yr$^{-1}(M_\mathrm{halo}/10^{12}M_\odot)$, where $M_\mathrm{halo}$ is the mass of the dark matter halo. At the inner edge, gas from the CGM can fall onto the galaxy, and feedback in the form of galactic winds are ejected (spherically) back out of the galaxy. The mass outflow rate of winds is related to the rate at which mass falls onto the galaxy (i.e. through the inner edge of the simulated domain) by
\begin{equation}
\dot M_\mathrm{out}=\frac{\eta}{\eta+1}\dot M_\mathrm{in},
\end{equation}
where $\eta$ is the mass-loading factor of the wind and is set to a constant value. The winds are also parameterized by the ejection velocity, $v_w$, which is set to a constant value proportional to the escape speed from the galaxy, $v_\mathrm{esc}$. At the inner edge of the domain, where the wind is launched, $v_\mathrm{esc}\approx3.5v_\mathrm{vir}$ where $v_\mathrm{vir}$ is the virial velocity of the halo. The wind is blowing for the full duration of the simulations.

The metallicity of the gas is fixed to one-third solar, and is the same for both the accreting gas and the galactic outflows. Outflows are expected to be metal-enriched compared to mostly-pristine cosmological accretion \citep{Muzahid2015,Chisholm2018,Christensen2018}, so this fixed metallicity is an idealized simplification, but aligns with some CGM studies that find a metallicity $\sim0.3 Z_\odot$ \citep{Prochaska2017,Muzahid2018}. The cooling rate of the gas is calculated assuming optically-thin photoionization equilibrium with a \citet{Haardt2001} ionizing background. The gas is allowed to radiatively cool, but there is a cooling floor of $T=10^4$ K, the temperature expected for photoionized gas, to prevent runaway unresolved cooling. Inside a core radius close to the galaxy, the gas is initially isentropic, and outside of that it is isothermal out to an initial virial shock \citep[see Figure 1 of][for the exact density, temperature, entropy, and cooling time initial conditions]{Fielding2017}. There are isobaric density perturbations throughout the simulation domain that break the spherical symmetry. These perturbations also cause the outflows to develop significant asymmetry. For example, if there is a particularly dense clump close to the inner boundary, winds injected in that direction will be immediately slowed by the dense clump while they continue escaping in other directions. The result is that winds tend to escape as ``jets" or ``fingers" with small opening angles and are highly asymmetric at any given time, but maintain symmetry when averaged over time. 

We focus on two simulations for the majority of this paper. The first, which has a larger dark matter halo mass, has $M_\mathrm{halo}=10^{12} M_\odot$, mass-loading $\eta=2$, wind speed $v_w^2=3v_\mathrm{esc}^2$, virial velocity $v_\mathrm{vir}=116$ km s$^{-1}$, and virial radius $R_\mathrm{vir}=319$ kpc \citep[the $10^{12}M_\odot$ simulation labeled ``fiducial high $\eta$" in Table 1 of][]{Fielding2017}. The inner edge of the domain is a sphere of radius 8 kpc, representing the galaxy. The outer edge of the domain is a sphere of radius $2R_\mathrm{vir}=638$ kpc. The second, lower-mass halo, has $M_\mathrm{halo}=10^{11}M_\odot$, mass-loading $\eta=5$, wind speed $v_w^2=3v_\mathrm{esc}^2$, virial velocity $v_\mathrm{vir}=54$ km s$^{-1}$, and virial radius $R_\mathrm{vir}=148$ kpc \citep[the $10^{11}M_\odot$ simulation labeled ``strong high $\eta$" in Table 1 of][]{Fielding2017}. The inner edge of this simulation domain is a sphere of radius $3.7$ kpc and the outer edge is a sphere of radius $2R_\mathrm{vir}=296$ kpc. The initial location of the virial shock is $0.25R_\mathrm{vir}$ for the lower-mass halo and $0.58R_\mathrm{vir}$ for the higher mass halo. Within this shock, the gas temperature is a constant $\sim4\times10^5$ K for the lower-mass halo and $\sim8\times10^5$ K for the higher-mass halo, and in both the gas temperature is $10^4$ K outside the shock. These wind parameters are representative of the observed winds from galaxies of these sizes \citep{Heckman2015,Muratov2015}.

The Eulerian nature and fixed cell size of these simulations (1.4 kpc and 0.65 kpc in the higher- and lower-mass halos, respectively, within $0.5R_\mathrm{vir}$, and 2.8 kpc and 1.3 kpc for the higher- and lower-mass halos for $0.5R_\mathrm{vir}<R<R_\mathrm{vir}$) allows for higher resolution to be reached in the diffuse CGM than in particle-based codes that focus on mass resolution rather than spatial resolution. This high resolution is crucial for disentangling the different phases and motions of gas present in the diffuse CGM \citep{Fielding2017,Hummels2018,Peeples2019}.

\begin{figure*}
\begin{minipage}{175mm}
\centering
\includegraphics[width=\linewidth]{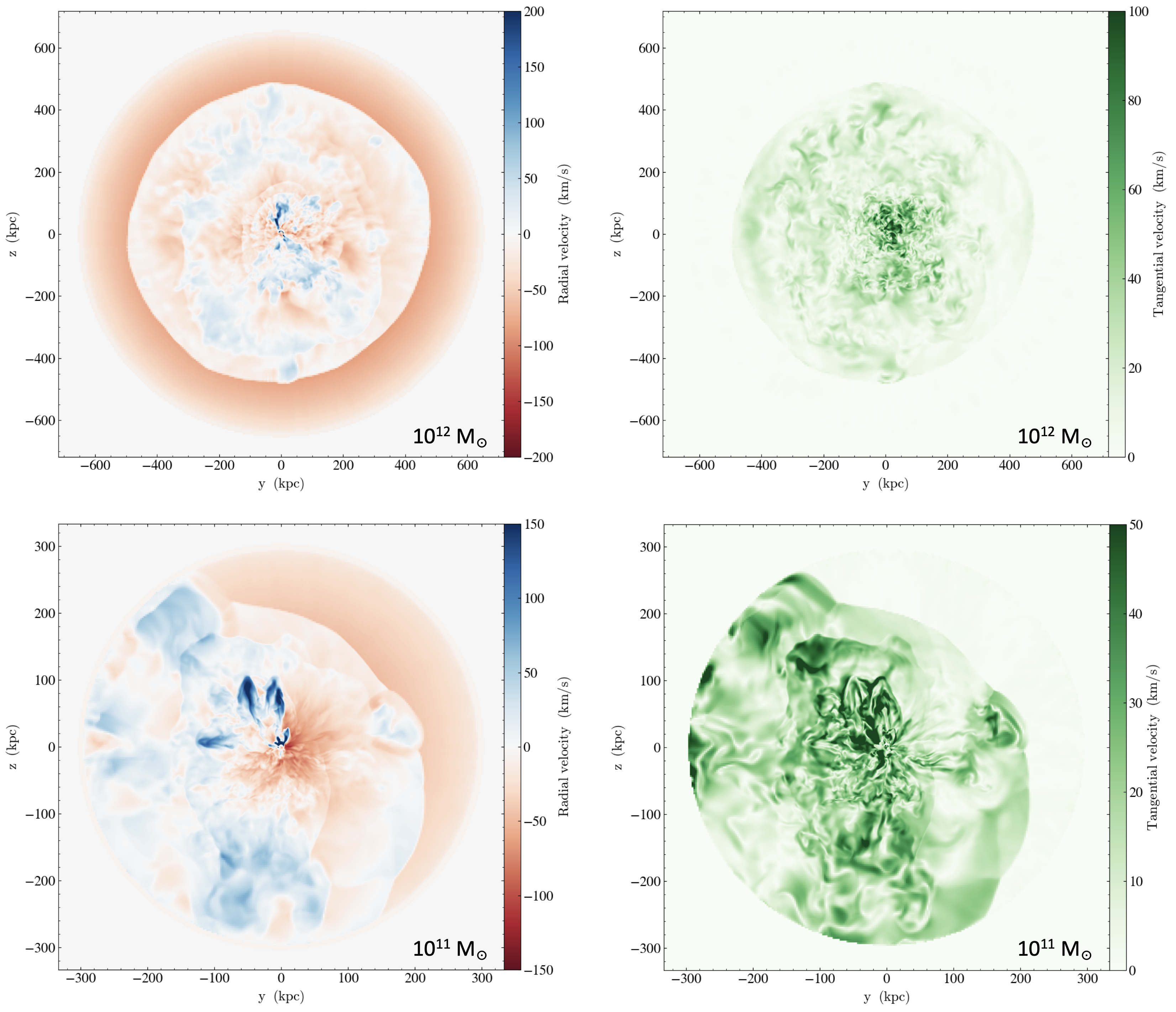}
\caption{Slices of radial velocity with respect to the galaxy at the center of the simulation domain (left column) and tangential velocity (right column) for both the higher-mass halo (top row) and lower-mass halo (bottom row), each at a time $\sim6$ Gyr into the simulation. Despite the spherically-symmetric initialization of the simulation, there are large-scale asymmetries that mimic the asymmetries of gas flows surrounding galaxies in fully cosmological simulations \citep[see][for more visualizations of these simulations]{Fielding2017}.}
\label{fig:v_slices}
\end{minipage}
\end{figure*}

The highly idealized nature of the simulations allow us to better understand the impacts of galactic outflows on the CGM. In a fully cosmological simulation, it is difficult to determine how each physical process (ouflows, cosmological inflow) contributes to the resultant CGM, especially in an Eulerian setup without tracer particles. By removing the complicating factor of cosmological filamentary accretion, we can more confidently attribute our results to the impact of galactic outflows. Although we make many simplifying assumptions, the simulations are calibrated to cosmological simulations \citep[for more details, see][]{Fielding2017}. For example, the wind outflow rates are those measured from the FIRE simulations in \citet{Muratov2015}. Figure~\ref{fig:v_slices} shows slices of radial velocity (relative to the galaxy at the center of the halo) and tangential velocity (velocities perpendicular to the radial velocity, see \S\ref{sec:velocities}) for both halos, after the simulations have been running for $\sim6$ Gyr. Despite the initial spherical symmetry, large asymmetries clearly develop, and outflows tend to be extended ``fingers" rather than spherically symmetric bubbles, as mentioned above. See \citet{Fielding2017} for more visualizations of these simulations.

\section{Results}
\label{sec:results}

To perform our analysis of the simulated CGM, we first chose 8 snapshots in time for each simulation, starting at $4.9$ Gyr (to avoid initial transients) and ending at $9.8$ Gyr, separated by $0.7$ Gyr. At each snapshot, we split the full simulation domain into 3D radial bins as fractions of the virial radius for each halo. Our CGM gas parameters of interest are the density, temperature, thermal pressure, spherical velocity components, and two components of dynamic pressure: in the radial direction and in the tangential directions. Using the python simulation analysis and visualization module \texttt{yt} \citep{Turk2011}, we computed a distribution (normalized by either mass or volume, depending on the parameter) of the values of parameters of interest of all cells that fall into each radial bin at that given time snapshot, then compute the median and interquartile range (IQR) representing the 25\% to 75\% interval of either mass or volume, depending on parameter, of all cells in the bin. We then averaged the distributions, medians, and IQRs over time and found the standard deviation of these values in time as well. In this way, the median value of a parameter in a radial bin represents the time-averaged value of the majority of the gas in the bin, and the IQR represents the time-averaged range of values of all gas in a given radial bin. The standard deviation on these parameters represents only the fluctuation in time.

Although the simulations are run for an extended period of time, we stress that there is no cosmic evolution or growth of the central galaxy included in these simulations. This simplification allows us to measure the properties of the CGM as they relate to exclusively the mass of the dark matter halo and the galactic outflows, without complicating effects such as galaxy environment, filamentary accretion, galaxy mergers, growth of galaxies over cosmic time, or any other complicating factors of galaxy evolution. Because there is nothing inherently evolving over the course of the simulations, each time snapshot (after initial transients) could be viewed as a different halo with the same mass. This allows us to obtain some CGM property statistics despite having just one simulation for each halo.

\subsection{CGM Pressure and Velocity Distributions}
\label{sec:hist}

\begin{figure*}
\begin{minipage}{175mm}
\centering
\includegraphics[width=\linewidth]{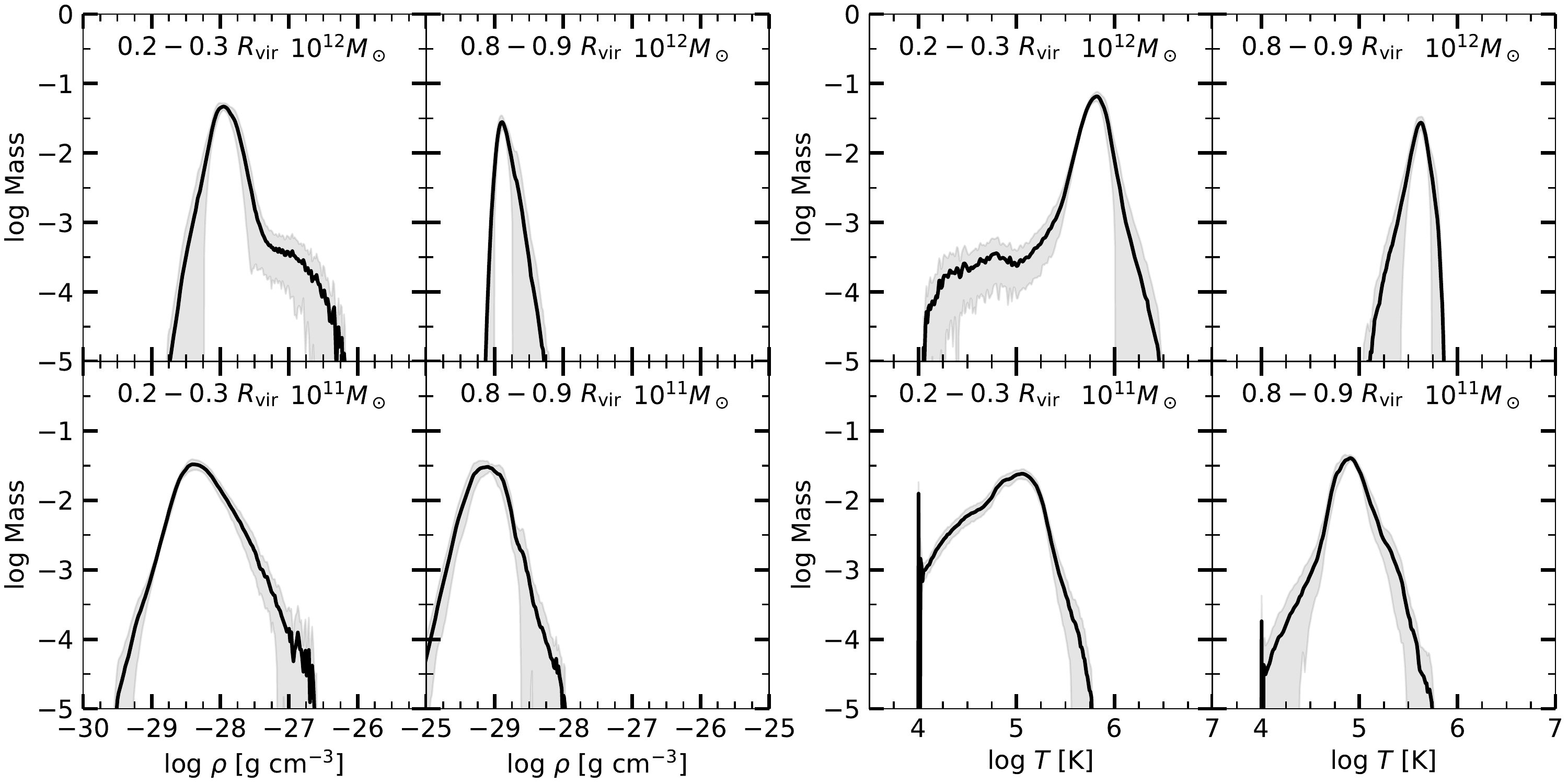}
\caption{Left 4-panel grouping shows distributions of the density within two radial bins, $0.2-0.3\ R_\mathrm{vir}$ (left) and $0.8-0.9\ R_\mathrm{vir}$ (second from left) for both the $10^{12}M_\odot$ (top) and $10^{11}M_\odot$ (bottom) halos. Right 4-panel grouping shows distributions of the temperature within the same radial bins. The distributions are averaged over the8 time snapshots linearly spread between $4.9$ Gyr and $9.8$ Gyr, and the shading shows one standard deviation of the distributions' variation with time. The distributions are not well-described by a single normal distribution.}
\label{fig:den_temp_hist}
\end{minipage}
\end{figure*}

It is well-known observationally that the CGM and galactic winds are both multiphase, with gas ranging from $\lesssim10^4$ K to $\gtrsim10^{6}$ K in temperature and similarly as many orders of magnitude in density \citep[see][for review]{Heckman2017,Tumlinson2017}. In addition, the peak of the radiative cooling curve near $10^5$ K allows warm gas to cool to $10^4$ K rapidly, so a multiphase medium is expected to form from thermal instability as cool gas precipitates out of the warm/hot medium. Figure~\ref{fig:den_temp_hist} shows the distributions of density and temperature within a small selection of radial bins for both the $10^{12}M_\odot$ and $10^{11}M_\odot$ halos. The multiphase nature of the simulated CGM is clearly evident --- even when considering gas all located at the same radius, the distributions of density or temperature are generally not well-described by a normal distribution, except perhaps for the density distributions in the lower-mass halo. Instead, we see that the bulk of the gas is located at a characteristic density or temperature, but there is a tail or second peak of the distribution toward low temperature or high density, especially in the radial bins close to the galaxy. Due to this non-symmetric shape of the distribution for density and temperature, especially in the higher-mass halo, the median and IQR are not good descriptors of the distribution. We direct the reader to \citet{Fielding2017} for more distributions of the density in radial bins (their Figure~10, see also Figure~7 for the density and temperature radial distributions) for these simulations.

\subsubsection{Thermal Pressures}

\begin{figure*}
\begin{minipage}{175mm}
\centering
\includegraphics[width=\linewidth]{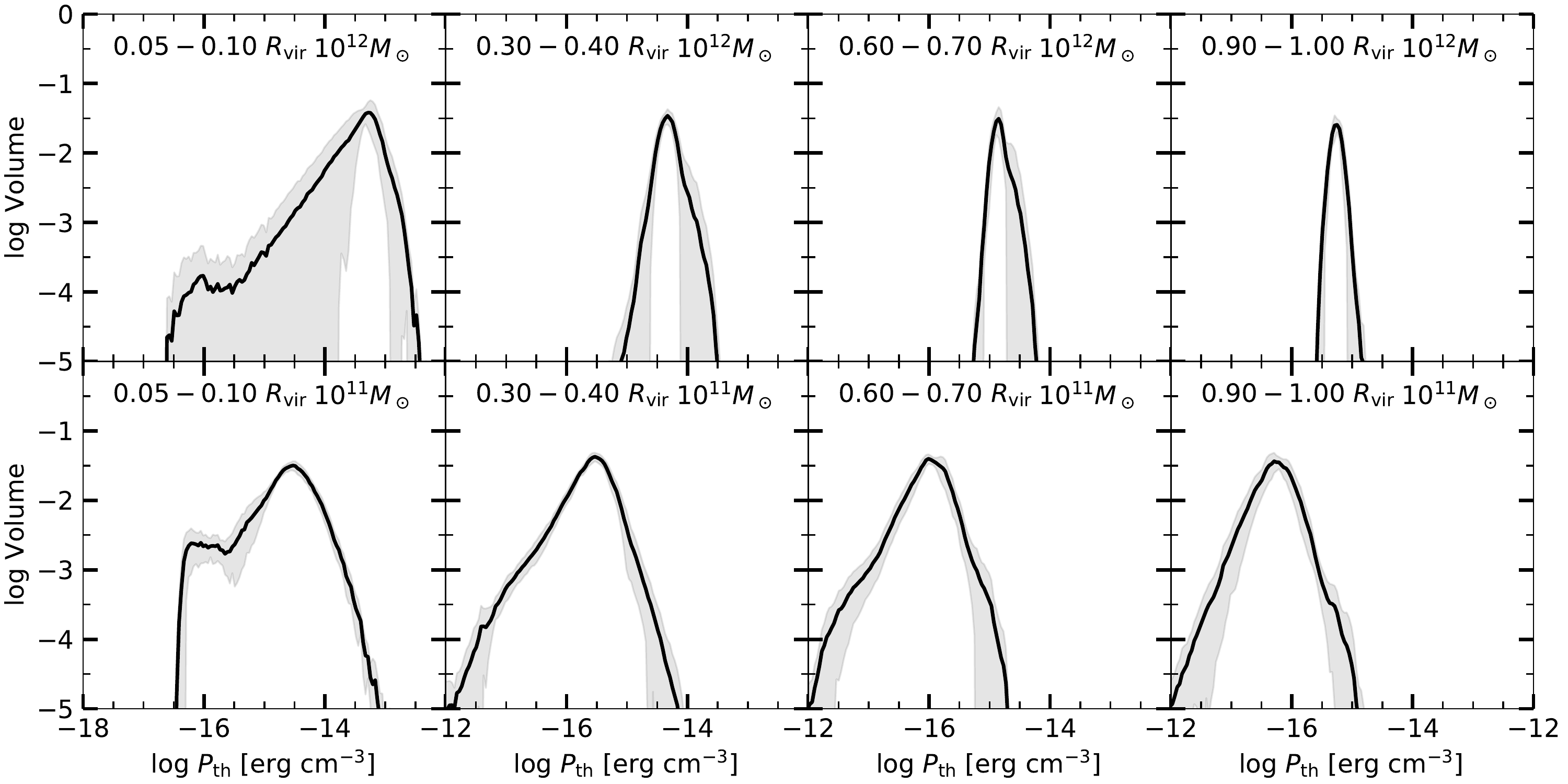}
\caption{Volume distributions of the thermal pressure in selected radial bins (columns) in the $M_\mathrm{halo}=10^{12}M_\odot$ (top row) and $M_\mathrm{halo}=10^{11}M_\odot$ (bottom row) simulations. The distributions are averaged over the 8 time snapshots linearly spread between $4.9$ Gyr and $9.8$ Gyr, and the shading shows one standard deviation of the distributions' variation with time. The pressure distributions are somewhat better described by a single normal distribution than the density or temperature distributions, especially in the higher-mass halo at intermediate to large radii.}
\label{fig:pres_hist}
\end{minipage}
\end{figure*}

The thermal pressure distributions are visually closer to normal distributions in radial bins far from the galaxy, so we characterize them with a median and IQR. Figure~\ref{fig:pres_hist} shows volume-weighted distributions of the thermal pressure, $P_\mathrm{th}$, in radial bins for both simulations.

In the smallest radius bin for both halos, the pressure distribution departs from a single log-normal and shows a tail to low pressure. The CGM close to the galaxy is split into low-pressure inflows and fast outflows, which causes the pressure distribution to depart from a normal distribution. The tail to low pressure is more time-variable than the primary peak (indicated by the shading in Figure~\ref{fig:pres_hist}) and is not ubiquitous at all times.

Compared to the larger-mass halo, the thermal pressure distributions in the lower-mass halo are wider, even if only the dominant peak of the distribution is considered. The lower-mass halo contains gas that has a wider range in thermal pressure than the higher-mass halo and is not as well-defined by a single gas pressure at each radius. In addition, the peak of the pressure distributions are located at a lower pressure than in the higher-mass halo, which is to be expected because the virial temperature of the lower-mass halo is significantly lower due to the shallower potential well, so it is initialized with a lower-temperature CGM. In both halos, the primary peak of the pressure distribution becomes narrower with increasing radial bin, as the halo gas becomes smoother further from the strong winds near the galaxy.

\subsubsection{Velocities}
\label{sec:velocities}

\begin{figure*}
\begin{minipage}{175mm}
\centering
\includegraphics[width=\linewidth]{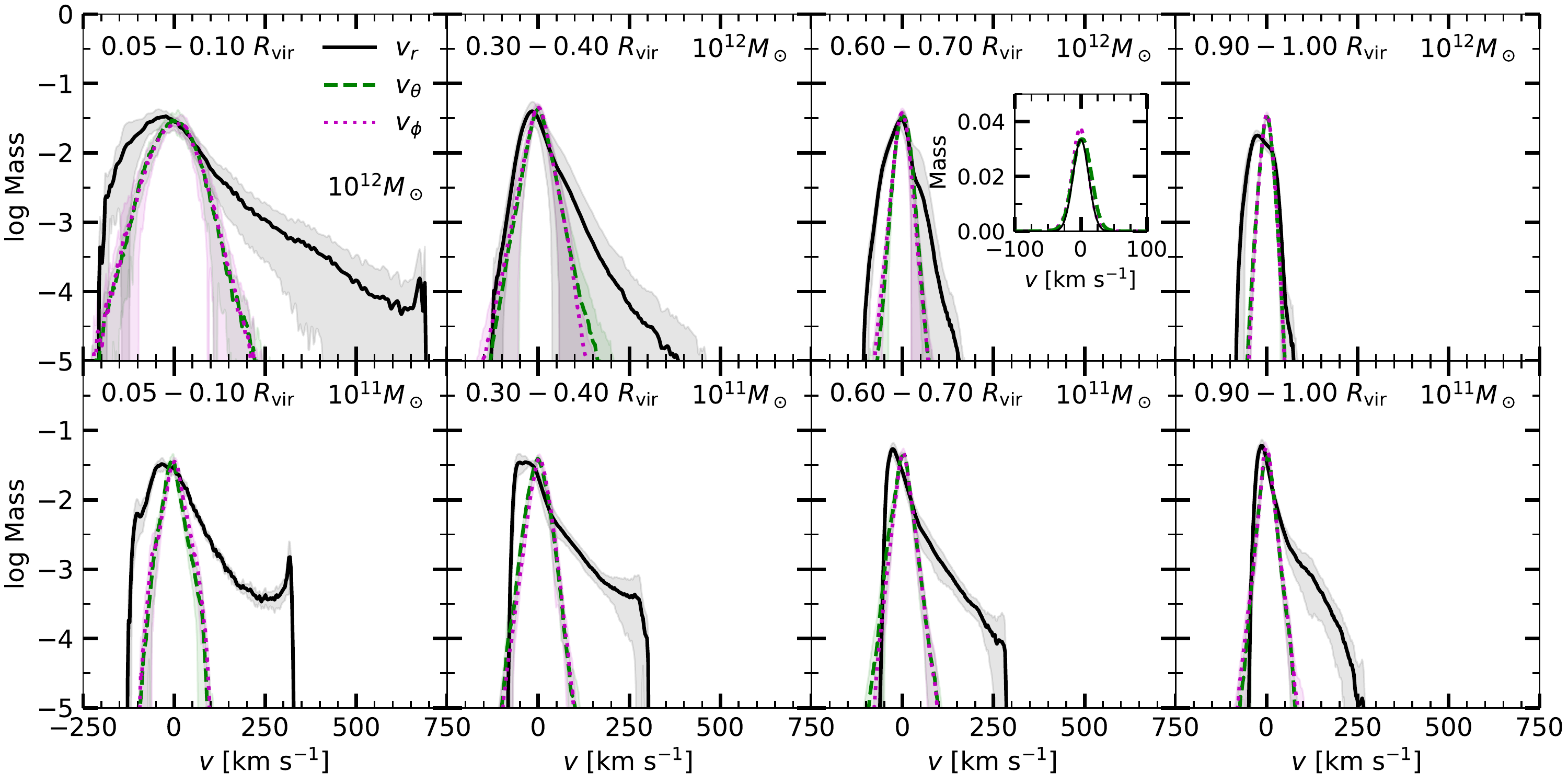}
\caption{Mass-weighted distributions of the three components of spherical velocity, radial velocity $v_r$ (black solid), $v_\theta$ (green dashed), and $v_\phi$ (magenta dotted) in radial bins (columns) in the $M_\mathrm{halo}=10^{12}M_\odot$ (top row) and $M_\mathrm{halo}=10^{11}M_\odot$ (bottom row) simulations. The distributions are averaged over the 8 time snapshots linearly spread between $4.9$ Gyr and $9.8$ Gyr, and the shading shows one standard deviation of the distributions' variation with time. The inset panel in the top, second from the right panel shows the $v_\theta$ and $v_\phi$ distributions in linear space as well as a normal distribution (thin black curve) with mean and standard deviation equivalent to that of the $v_\theta$ and $v_\phi$ distributions. The normal distribution sits nearly on top of the $v_\theta$ and $v_\phi$ distributions, indicating they are well-described by a normal distribution and therefore possibly tracing isotropic turbulence, unlike the $v_r$ distributions that trace bulk flows}.
\label{fig:vel_hist}
\end{minipage}
\end{figure*}

We split the velocity of the gas into the 3D spherical velocity components: radial velocity $v_r$, velocity in the $\theta$ direction $v_\theta$, and velocity in the $\phi$ direction $v_\phi$. These simulations do not have bulk rotation, so we do not expect any systematic differences between $v_\theta$ and $v_\phi$. However, due to the radial outflows and accretion, we do expect $v_r$ to vary significantly from either of the other velocity directions. We plot the mass-weighted distributions of the three components of velocity (different from the volume weighting of the pressure distributions) within radial bins in Figure~\ref{fig:vel_hist}.

The $v_\theta$ and $v_\phi$ distributions are essentially equivalent to each other in all radial bins, as expected. The inset in the $0.6-0.7\ R_\mathrm{vir},\ 10^{12}M_\odot$ panel shows the $v_\theta$ and $v_\phi$ distributions in linear space, where it is clear that they are well-described as normal distributions with no significant tails to either high or low velocity. A thin black curve in this inset panel shows a normal distribution with mean and standard deviation equivalent to that of the $v_\theta$ and $v_\phi$ distributions, and it lies nearly on top of the $v_\theta$ and $v_\phi$ distributions. Because the velocities in these directions do not contain any information about outflows or inflows (the distributions are peaked at 0 km s$^{-1}$ in every radial bin, and strongly time-invariable), and because they are equivalent to each other, we argue they trace the turbulent velocity in each radial bin. The $v_\theta$ and $v_\phi$ distributions become narrower with increasing galactocentric radius, indicating that turbulent velocities are higher in the inner regions of the halo.

The $v_r$ distribution is very different. This is expected, as the $v_r$ component of the velocity includes the radial outflows and inflows. The outflow is evident in the radial velocity distribution as a long and time-variable tail to positive velocities, which extends to higher velocities at smaller radii. At all radii, there is also a signature of inflow that appears as a small ``bump" or a second sharp peak in the distribution at negative velocities, or manifests as the distribution being peaked at a negative velocity. The equivalent distributions of $v_\theta$ and $v_\phi$ centered at 0 km s$^{-1}$ show the turbulence is isotropic in those directions, so the $v_r$ distribution may also contain a component of turbulence in the radial direction. However, this is difficult to separate from the bulk flows. We see that there is a signature of inflow in all radial bins, not just the smallest, which is evidenced by the small excess of $v_r$ over $v_\theta$ or $v_\phi$ at negative velocities.

In the lower-mass halo, the width of both the radial and tangential distributions are narrower, which is expected if the bulk radial flows seed the turbulence because the outflows have a lower velocity by design than the $10^{12}M_\odot$ halo, as its virial velocity is lower. The fastest-flowing gas in each halo is roughly 4.5 times the virial velocity of each halo, so the differences in velocity distributions between the halos can be attributed to the differences in their virial velocities more than any difference in the driving of the winds of turbulent motions themselves. The $v_\theta$ and $v_\phi$ distributions do not narrow with increasing radius as strongly in the lower-mass halo as in the higher-mass halo. This halo is somewhat more inflow-dominated than the higher-mass halo. The second, very narrow peak in the $v_r$ distribution at negative velocities indicates the constant-velocity cosmological accretion present in both halos. In the higher-mass halo, the accretion is only present at large galactocentric radii, but in the lower-mass halo, it can be seen down to $0.3-0.4R_\mathrm{vir}$. The lower-mass halo develops significant asymmetries in the outflow, so smooth cosmological accretion may reach small radii in some directions while it is disrupted by outflows in other directions, producing the bimodal shape of the $v_r$ distribution in most radius bins.

\subsection{Radial Trends}
\label{sec:pres}

To examine how the global properties of the CGM change with galactocentric radius, we compute the median and interquartile range of the middle 50\% of the volume-weighted distributions of thermal pressure and of the mass-weighted distributions of radial and tangential velocities, and then average these over the 8 time snapshots. We compute this within radial bins of width $0.1R_\mathrm{vir}$ for each halo. We plot all radii in units of kpc for easy physical dimension comparison, but note that the points on each halo's curve in the subsequent plots represent each of the 10 radius bins in fractions of the virial radius.

We define two types of effective dynamic pressure\footnote{We define ``dynamic" pressure analogously to ram pressure of a flow: $P=\rho v^2$, but consider different directions of $v$ to distinguish between bulk flows and turbulence.}: the radial dynamic pressure, $P_\mathrm{rad}=\rho v_r^2$, and the tangential dynamic pressure, $P_\mathrm{tan}=\rho v_\mathrm{tan}^2$, where the tangential velocity, $v_\mathrm{tan}^2=\frac{1}{2}(v_\theta^2+v_\phi^2)$, is the average in quadrature of the $\theta$ and $\phi$ direction velocities. $P_\mathrm{tan}$ can be thought of as tracing the turbulent pressure of the gas in the simulations, as the tangential velocity distributions within radial bins are isotropic and appear to be tracing turbulence, as discussed above. $P_\mathrm{rad}$ can be thought of as tracing the ram pressure of the inflows and outflows in the simulations, but it contains some contamination of turbulent motions in the radial direction, so it is an upper limit on ram pressure from purely bulk flows in and out. The true, total dynamic pressure due to both turbulence and ram pressure of bulk flows lies somewhere between $P_\mathrm{tan}$ and $P_\mathrm{rad}$, but it is impossible to determine how much of a given pixel's velocity is contributed by the radial velocity distribution and how much by the tangential velocity distribution so we cannot separate the tangential pressure from the radial pressure any further. In order to separate the contribution of ram pressure due to inflows from that due to outflows, we also calculate $P_\mathrm{rad,in}$ and $P_\mathrm{rad,out}$, where only gas with negative radial velocities contribute to the former and only gas with positive radial velocities contribute to the latter. Both $P_\mathrm{rad,in}$ and $P_\mathrm{rad,out}$ contain contributions from the isotropic turbulent pressure in addition to the ram pressure of the bulk flows, but because we expect the turbulence to be isotropic, the turbulent pressure should contribute equally to both.

We find the mean value of these dynamic pressures within each radial bin and compute the standard deviation of the fluctuation of the mean across the different time snapshots. We also use these definitions of dynamic pressure to compute the overall pressure support of the CGM in \S\ref{sec:HSE}.

\subsubsection{Pressures}

\begin{figure*}
\begin{minipage}{175mm}
\centering
\includegraphics[width=\linewidth]{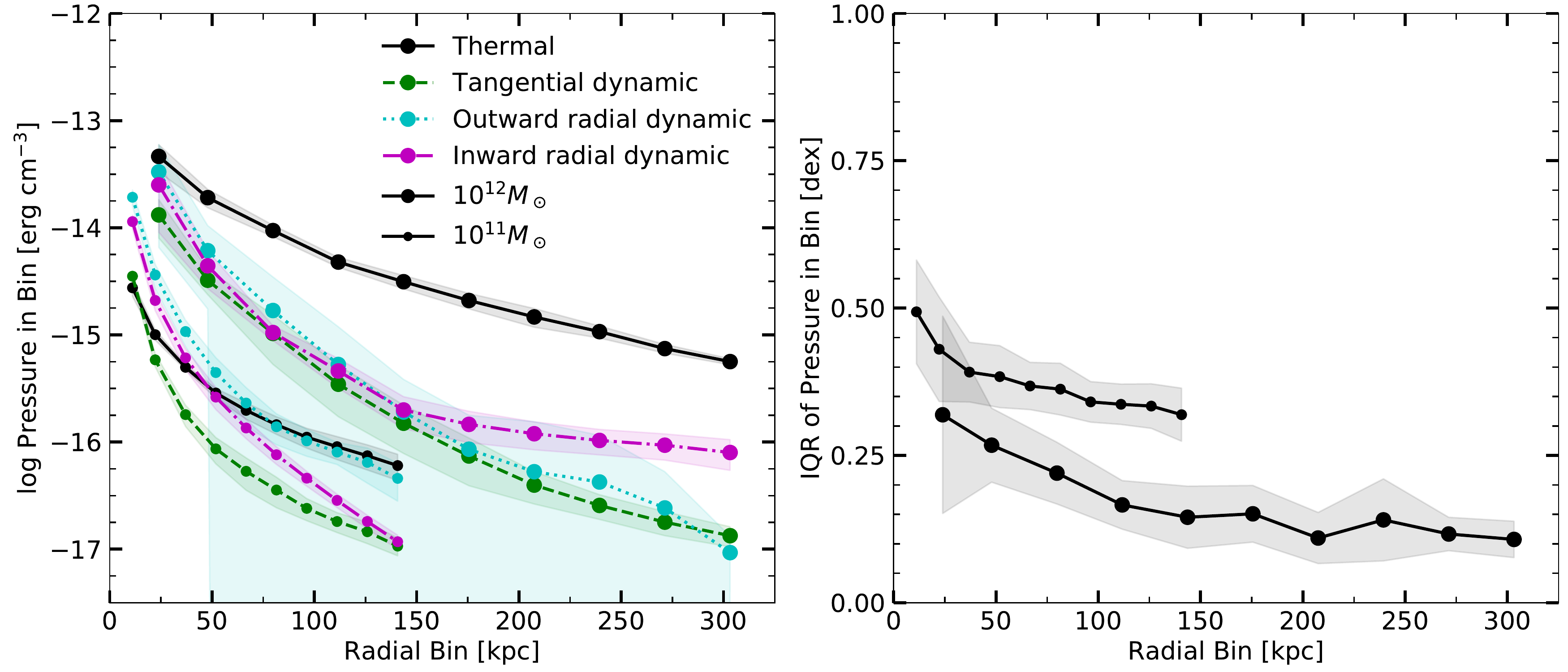}
\caption{Left panel shows as functions of radius the time-averaged median of the thermal pressure (black symbols and solid lines) and the time-averaged mean of the dynamic pressures: tangential (green symbols and dashed lines), outward radial (cyan symbols and dotted lines), and inward radial (magenta symbols and dot-dashed lines), in radial bins, for both the $10^{12}M_\odot$ (large circles) and $10^{11}M_\odot$ (small circles) halos. Right panel shows the time-averaged thermal pressure IQR in both halos. Shaded regions around curves indicate the one standard deviation time variance of the values. Thermal pressure is most important in the higher-mass halo while the effective pressure of bulk flows are equally as important as thermal pressure in the lower-mass halo.}
\label{fig:pres_radius}
\end{minipage}
\end{figure*}

Figure~\ref{fig:pres_radius} shows the median and IQR of the thermal pressure distribution, the mean outward radial dynamic pressure, the mean inward radial dynamic pressure, and the mean tangential dynamic pressure, computed within radial bins, as functions of radius. Both halos show smoothly decreasing median thermal and mean dynamic pressures with increasing radius, but all types of pressure in the higher-mass halo have larger values than all pressures in the lower-mass halo. For the higher-mass halo, the thermal pressure is always larger than the dynamic pressures. In the inner regions of the higher-mass halo, the thermal and dynamic pressures are much closer in value than in the outskirts of the halo, where the thermal pressure can be $1-1.5$ orders of magnitude larger than the dynamic pressures. In the lower-mass halo, the inner regions of the halo are dominated by radial dynamic pressure, likely due to the fast and strong inflows and outflows in this halo, while the thermal pressure and outward radial dynamic pressures are roughly equal and dominant over the tangential and inward radial dynamic pressures in the outskirts of the halo. The similarity between the tangential and radial dynamic pressures in the higher-mass halo may indicate that this halo mass is more efficient at converting bulk radial flows into isotropic turbulence than the lower-mass halo, in which the tangential dynamic pressure is always significantly smaller than the radial dynamic pressures. However, the radial dynamic pressures have a large contribution from the radial direction of the isotropic turbulent motions, so we cannot exactly compute the efficiency of conversion from bulk flows to isotropic turbulence.

The IQR of the thermal pressure in the higher-mass halo is smaller, indicating a narrower distribution in a given radial bin, than the lower-mass halo. The time variation of the thermal pressure IQR is similar in both halos.

Altogether, Figure~\ref{fig:pres_radius} indicates that the higher-mass halo's thermal pressure is very well-behaved: it varies smoothly with galactocentric radius and has small IQRs. The thermal pressure in the lower-mass halo, by contrast, has a larger IQR (although it is also smoothly varying with galactocentric radius), possibly indicating that the lower-mass halo is not as thermal pressure-regulated as the higher-mass halo. Not only is the higher-mass halo's thermal pressure well-behaved, it also dominates over other forms of pressure in this halo. In the lower-mass halo, bulk inflows and outflows are a stronger contributor to the overall pressure than in the higher-mass halo.

\begin{figure*}
\begin{minipage}{175mm}
\centering
\includegraphics[width=\linewidth]{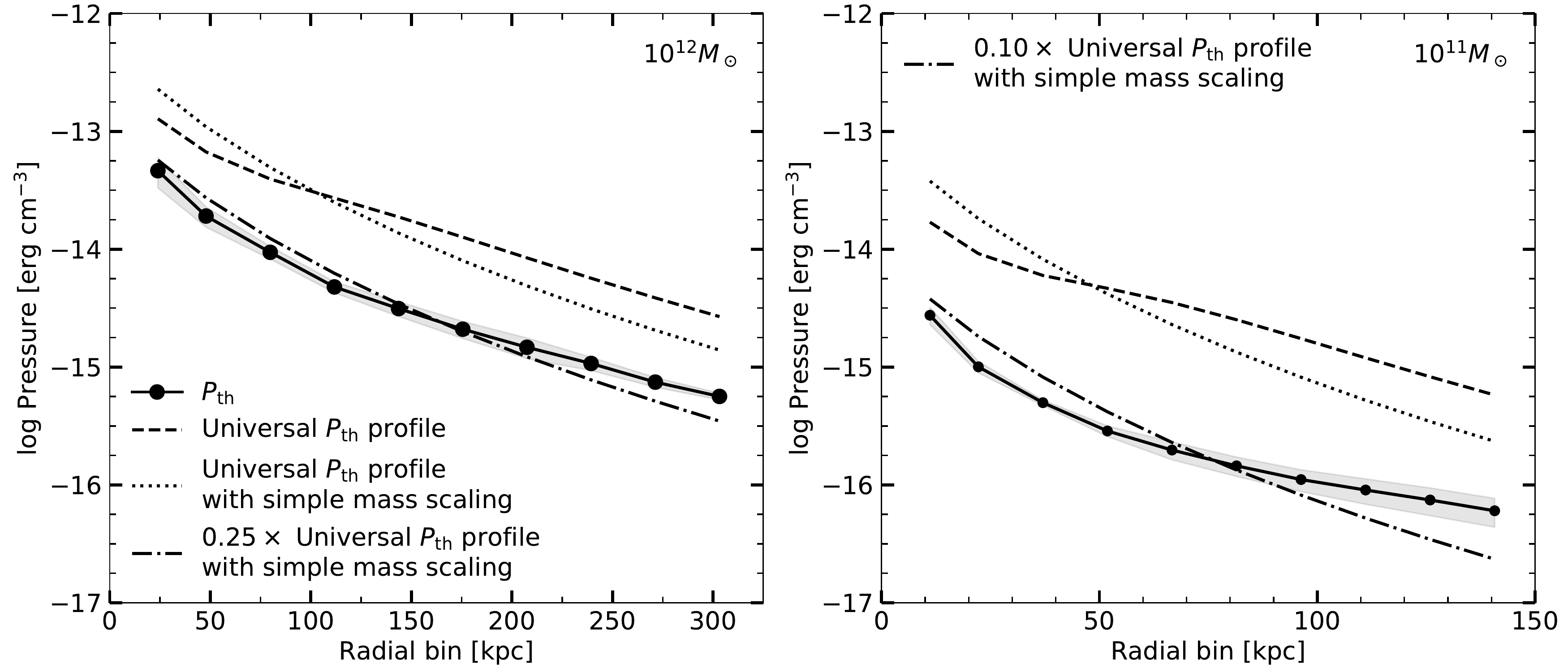}
\caption{The thermal pressure (black circles) as a function of radius in the $10^{12}M_\odot$ halo (left) and the $10^{11}M_\odot$ halo (right) compared to three forms of the universal pressure profile as given by \citet{Arnaud2010}: the full form of the universal pressure profile in which the slope of the mass dependence is also radius-dependent (dashed), a simpler form where the slope of the mass dependence is constant (dotted), and the simpler form renormalized by an arbitrary factor in an attempt to better match the pressure profiles in our simulations (dash-dotted). The renormalization factor is $0.25$ in the higher-mass halo and $0.1$ in the lower-mass halo. The thermal pressure profiles in these halos do not seem to be well-described by the empirical universal pressure profile for galaxy clusters.}
\label{fig:univ_pressure}
\end{minipage}
\end{figure*}

To determine if the pressure variation with radius in these idealized simulations matches the ``universal pressure profile" found for hot gas in galaxy clusters \citep{Arnaud2010,Planelles2017}, we compared the pressure profile empirically determined by \citet{Arnaud2010} to the thermal pressure profiles in our simulated CGM in Figure~\ref{fig:univ_pressure}, rescaled to the appropriate halo mass. We found that neither the normalization nor the slope of the pressure profiles in the idealized simulations studied here matches the universal pressure profile; the slope we find is steeper and the overall normalization lower. In the universal pressure profile, the mass of the halo is raised to a power that depends on the radius within the halo \citep[see equations~8 and~10 of][]{Arnaud2010}. A simpler form of the universal pressure profile sets this power to a constant, without any radial dependence, so we try both these forms of the universal pressure profile. We additionally scale the overall normalization of this simpler form by arbitrary values to attempt to match our simulated pressure profiles in the dash-dot curves. The simpler, scaled form of the universal pressure profile nearly matches the pressure profile we find in the higher-mass halo, but there is no physical reasoning for the normalization scaling. In the lower-mass halo, even the simpler, scaled form of the universal pressure profile does not match our pressure profile well. Perhaps the universal pressure profile does not maintain its shape and normalization when scaled down to the mass of single-galaxy hosting halos, or perhaps these idealized simulations are not capturing a physical process present in real galaxy cluster gaseous halos.

\subsubsection{Radial Velocities}
\label{sec:vel}

\begin{figure*}
\begin{minipage}{175mm}
\centering
\includegraphics[width=\linewidth]{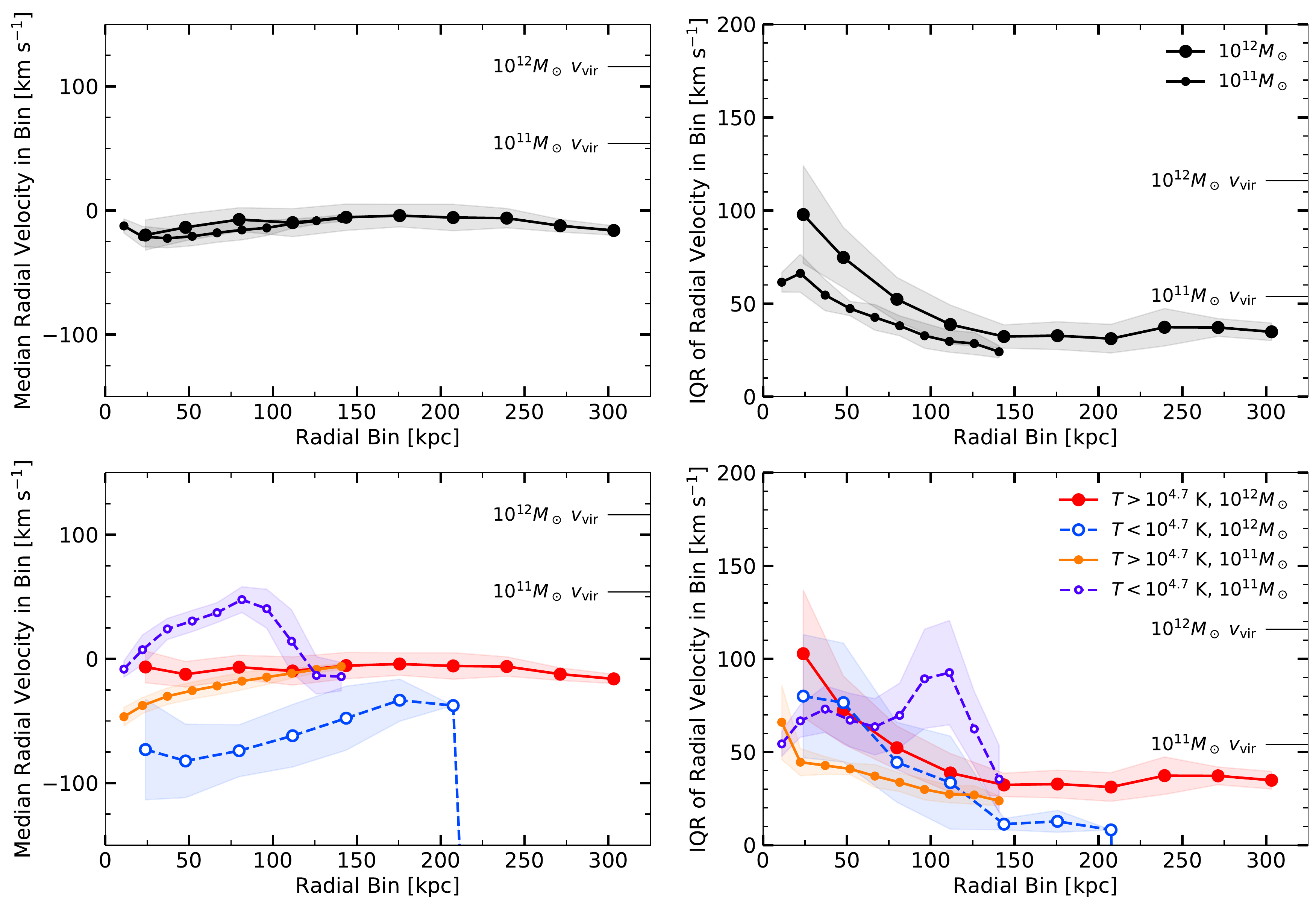}
\caption{Left panels show the time-averaged median radial velocity as a function of radius for both the $10^{12}M_\odot$ (circles) and $10^{11}M_\odot$ halos. Right panels show the time-averaged IQR of the radial velocity distributions in both halos. Top panels show radial velocities for all gas in each simulation, while bottom panels show the median and IQR of the radial velocity distributions when the gas is first separated by temperature at $T=10^{4.7}$ K. Shaded regions around curves indicate the one standard deviation time variance of the values. Note that the values in the top panels are the averages of the bottom panels only when weighted by the mass of gas in each temperature phase. The value of each halo's virial velocity is plotted as a black solid line with a label on the right axis of each panel for comparison. The high-mass halo has static hot gas and inflowing cold gas while the low-mass halo has inflowing hot gas and outflowing cold gas.}
\label{fig:vr_radius}
\end{minipage}
\end{figure*}

The strong contribution of dynamic pressures to the overall pressure of the CGM, especially in the lower-mass halo, warrant a deeper discussion of the velocity distributions. Figure~\ref{fig:vr_radius} shows the time-averaged median and IQR of the radial velocity distributions in each radial bin, for both halos. The virial velocity of both halos is marked on the right axis in each panel for comparison; $10^{12}M_\odot\ v_\mathrm{vir}=116$ km/s and $10^{11}M_\odot\ v_\mathrm{vir}=54$ km/s.

Considering the top-left panel of Figure~\ref{fig:vr_radius} first, both halos are inflow-dominated because the median of the radial velocity distributions in each radial bin is negative, as we saw previously in Figure~\ref{fig:vel_hist}, where the peaks of the radial velocity distributions were negative.

The top-right panel of Figure~\ref{fig:vr_radius} shows that the higher-mass halo has a slightly wider spread of radial velocities at each radius. However, when the velocities are normalized by each halo's virial velocity, the lower-mass halo has a wider spread of velocities, despite the fact that the outflow speed relative to the escape velocity is equivalent in both halos, indicating that the gas motions in the lower-mass halo are more anisotropic than in the higher-mass halo. For both halos, the width of the radial velocity distribution decreases with increasing radius because the outflows become smoother and more isotropic as they expand to larger distances from the galaxy.

The bottom panels of Figure~\ref{fig:vr_radius} show the median and IQR of radial velocity distributions containing only the hot gas at $T>10^{4.7}$ K or only the cold gas at $T<10^{4.7}$ K. We choose $10^{4.7}$ K as the dividing temperature because this temperature separates the bimodal phase distributions of all gas in the simulations well for both the higher-mass and the lower-mass halos (see Figure~8 in \citet{Fielding2017} for the temperature of all mass in the simulations over time). Note, however, that the gas with temperature above $10^{4.7}$ K in the higher-mass halo is at a higher temperature than the gas above this dividing line in the lower-mass halo, and the division between the two gas temperatures is more distinct in the higher-mass halo than in the lower-mass halo, especially at later times in the simulations.

The similarities between the halos vanish when the gas is split into high-temperature and low-temperature. Considering the lower-left panel of Figure~\ref{fig:vr_radius}, the median radial velocities of the gas above and below the dividing temperature in the higher-mass halo are clearly separated. The hot gas is still dominated by inflow, but at much slower inflowing velocities $\sim-10$ km s$^{-1}$ than the cold gas, which has inflowing velocities $\sim-50 - -100$ km s$^{-1}$. In the lower-mass halo, the median radial velocities of the hot and cold gases are flipped such that the cold gas has a positive or closer to zero median velocity, indicating outflow, in most radial bins, while the hot gas is more quickly inflowing in most radial bins. Unlike in the higher-mass halo, the lower-mass halo's hot and cold gas median velocities cross each other at large radius, where the hot gas represents the largest extent of outflows and the cool gas represents cosmological accretion. For both halos, the hot gas radial velocity is less time-variable than the cold gas. The bottom-right panel of Figure~\ref{fig:vr_radius} shows that the outflows that widen the radial velocity distributions in a given radial bin are primarily hot gas in the higher-mass halo, but primarily cold gas in the lower-mass halo. This is because a higher mass loading factor for the wind leads to more efficient radiative cooling \citep{Thompson2016}, and the lower-mass halo's wind is initialized with a higher mass loading factor than the higher-mass halo.

In summary, in the high-mass halo the hot gas is fairly static while the cool gas is inflowing, and in the low-mass halo, neither the hot nor the cold gas is static, with cold gas primarily outflowing (except at the outer edges of the halo where it is inflowing due to cosmological accretion) and hot gas primarily inflowing. This suggests that the source of cool gas in the high-mass halo may be cooling condensation due to thermal instability of a static hot halo. In the low-mass halo, the cool gas is produced through adiabatic and radiative cooling of the winds themselves, which is much more anisotropic.

\subsubsection{Tangential Velocities}
\label{sec:turb}

\begin{figure}
\centering
\includegraphics[width=\linewidth]{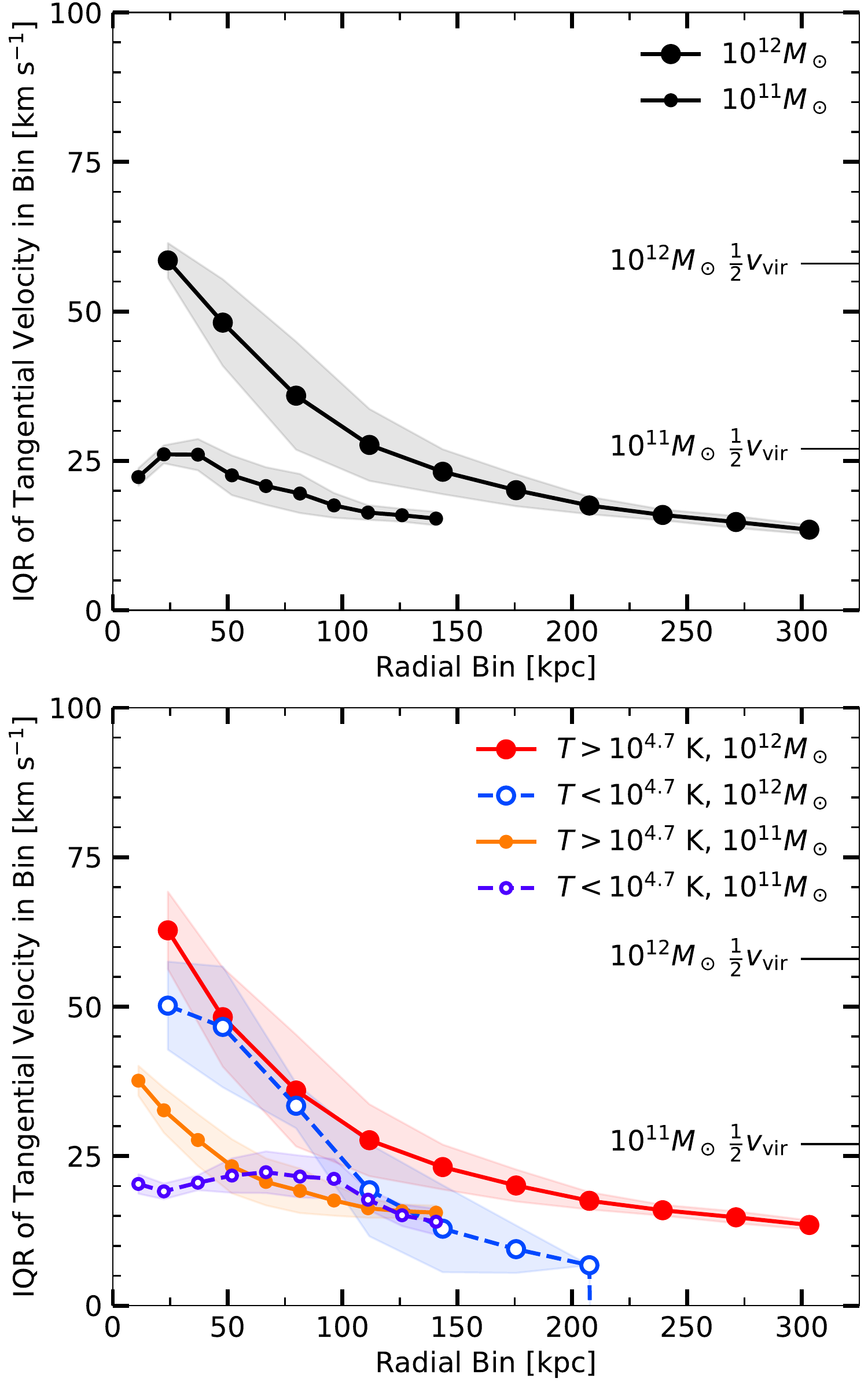}
\caption{The time-averaged IQR of the tangential velocity as a function of radius for both the $10^{12}M_\odot$ (large circles) and $10^{11}M_\odot$ (small circles) halos. Top panel shows tangential velocities for all gas in each simulation, while bottom panel shows the IQR of the tangential velocity distributions when the gas is first separated by temperature at $T=10^{4.7}$ K (hot: red and orange filled symbols; cold: blue and purple open symbols). Shaded regions around curves indicate the one standard deviation time variance of the values. Note that the values in the top panel are the averages of the bottom panel only when weighted by the mass of gas in each temperature phase. Half of the virial velocity value is labeled on the right axis in both panels for comparison. Both halos show similar turbulent velocities when normalized by the virial velocity of the halo, but the higher-mass halo has comparatively lower turbulence at large radii.}
\label{fig:vt_radius}
\end{figure}

Figure~\ref{fig:vt_radius} shows the time-averaged IQRs of the tangential velocity as a function of galactocentric radius, for all gas in the top panel and with gas split into hot ($T>10^{4.7}$ K) and cold ($T<10^{4.7}$ K) in the bottom panel. Because the medians of the $v_\theta$ and $v_\phi$ distributions are zero in all radial bins (see Figure~\ref{fig:vel_hist}), we show only the IQRs of the tangential velocity.

The higher-mass halo has a wider tangential velocity distribution in a given radial bin than the lower-mass halo at all radii, so it has a faster turbulent velocity. The two halos have roughly equivalent turbulent velocities when normalized by virial velocity; the maximum IQR, which occurs at $\sim25$ kpc from the galaxy, is $\sim0.5v_\mathrm{vir}$ for both halos. Both halos' tangential velocities decrease with increasing galactocentric radius, indicating that the turbulent velocity is higher in the central regions of the halo due to the strong galactic winds. The higher-mass halo has lower turbulent velocities relative to its virial velocity in the outskirts of the halo than the lower-mass halo, indicating the lower-mass halo maintains a high level of turbulence even at large galactocentric radii. The bottom panel of Figure~\ref{fig:vt_radius} shows that in the higher-mass halo, the hot gas in the higher-mass halo maintains a higher level of turbulence in the outer regions of the halo than the cold gas. In the lower-mass halo, the cold gas has much lower turbulent speeds in the innermost radial bins than the hot gas, but the turbulent velocities in both the hot and cold gas are similar at larger radii. The similar turbulent velocities at both high and low temperatures could be due to the fact that the ``hot" and ``cold" gases are actually quite close in temperature in the lower-mass halo.

\begin{figure*}
\begin{minipage}{175mm}
\centering
\includegraphics[width=\linewidth]{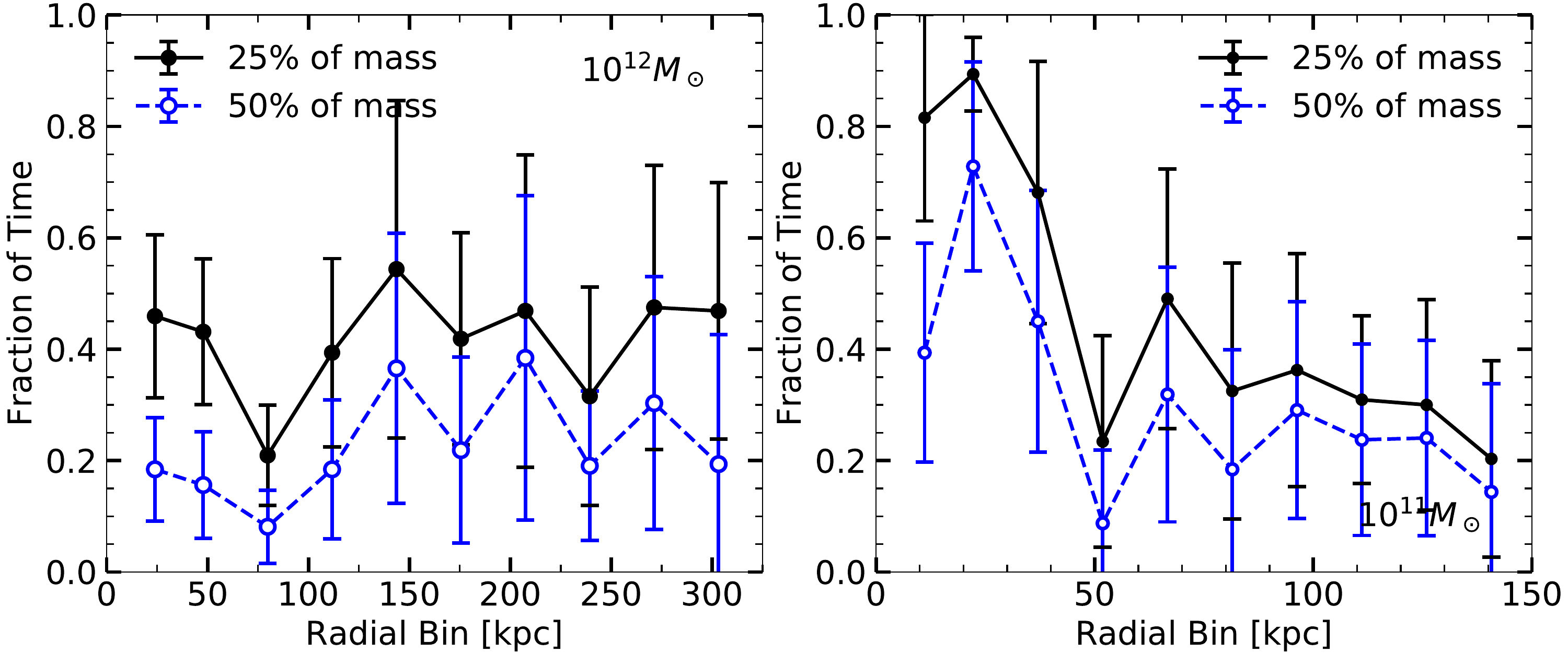}
\caption{The fraction of time that 25\% (filled points) or 50\% (open points) of the mass in a patch of the CGM has a radial velocity outside of the 95\% width of its tangential velocity distribution, averaged over 26 (possibly overlapping) patches within each radial bin, as a function of radius. Error bars show one standard deviation of this fraction across all patches within each radial bin. Left panel shows the higher-mass halo and right panel shows the lower-mass halo. A larger time fraction on the vertical axis indicates that the CGM, at that radius, is more dominated by bulk flows. A smaller fraction indcates that the CGM at that radius is more dominated by turbulent motions. Large error bars indicate a wide variety of radial velocity distributions at that radius. The inner regions of the low-mass halo spend more time dominated by bulk flows (rather than turbulence) than in the high-mass halo.}
\label{fig:time_fraction}
\end{minipage}
\end{figure*}

The time variance in Figure~\ref{fig:vel_hist} suggests that there is a non-negligible fraction of the time snapshots we explore in which the radial velocity distribution is similar to the tangential velocity distribution, suggesting that there is some amount of time when radial bulk flows are absent and the radial velocity is dominated by isotropic turbulence. To explore what fraction of the time a given part of the CGM is dominated by radial bulk flows vs. turbulence, we use the same 8 time snapshots as before and investigate radial velocity distributions in ``patches" of the CGM. We define 26 spheres in each radial bin, with diameters equal to the width of the radial bin, evenly spread in 3D space throughout the radial shell. Note that this definition can produce overlapping spheres, especially in the inner halo, as some spheres are separated by distances less than their radii. We first define a $2\sigma$ width of the tangential velocity distribution for all spheres in each radial bin, then calculate the fraction of each sphere's mass that has a radial velocity that falls outside this $2\sigma$ range. We do this for every time snapshot, then find the fraction of the 8 time snapshots where greater than 25\% (or 50\%) of a sphere's mass has a radial velocity outside of the $2\sigma$ range. A fraction of 0 indicates a patch of the CGM at that radius is always dominated by turbulence, while a fraction of 1 indicates it is always dominated by bulk flows. Note that because we use only 8 time snapshots, the time fractions can only take discrete values. We then average this time fraction across all spheres in each radial bin, and plot this average and standard deviation of time fractions as a function of radius in Figure~\ref{fig:time_fraction}.

As a function of radius, the higher-mass halo shows a roughly constant fraction of time when the CGM is dominated by bulk flows rather than turbulent motions, around $30\%$ of the time. In this case, the interaction of galactic outflows and cosmological accretion inflows primarily triggers turbulent motions in the halo rather than maintaining bulk motions.

The lower-mass halo shows both overall larger time fractions when the CGM is dominated by bulk flows at small radii, as well as a radial trend: the inner regions of the halo are more dominated by bulk motions (time fractions for 50\% of the mass $>50\%$) while the outer regions are similar to the higher-mass halo, with $70-80\%$ of their time dominated by turbulent motions. This is consistent with Figure~\ref{fig:pres_radius} where the radial dynamic pressure in the lower-mass halo is both more important than the turbulent pressure in the lower-mass halo and a larger contributor to the overall pressure of the halo than in the thermal pressure-dominated higher-mass halo.

\subsubsection{Velocity Structure Function}

In order to understand the typical length scale of coherent gas motions in the halo, if there is one, we compute the velocity structure function (VSF), which describes the difference in velocity vectors between two points in the halo. The shape and slope of the VSF tracks the type of turbulence dissipation and scale on which there are turbulent motions, or the ``driving scale" of the turbulence. Kolmogorov turbulence \citep{Kolmogorov1941} predicts a power-law slope of the VSF of $2/3$, which is also observed for turbulent clouds in the interstellar medium \citep{Larson1981}. The largest physical scale where the $2/3$ slope is measured indicates the starting point of turbulent cascades, or the scale on which the turbulence is being driven \citep{Li2019}.

\begin{figure*}
\begin{minipage}{175mm}
\centering
\includegraphics[width=\linewidth]{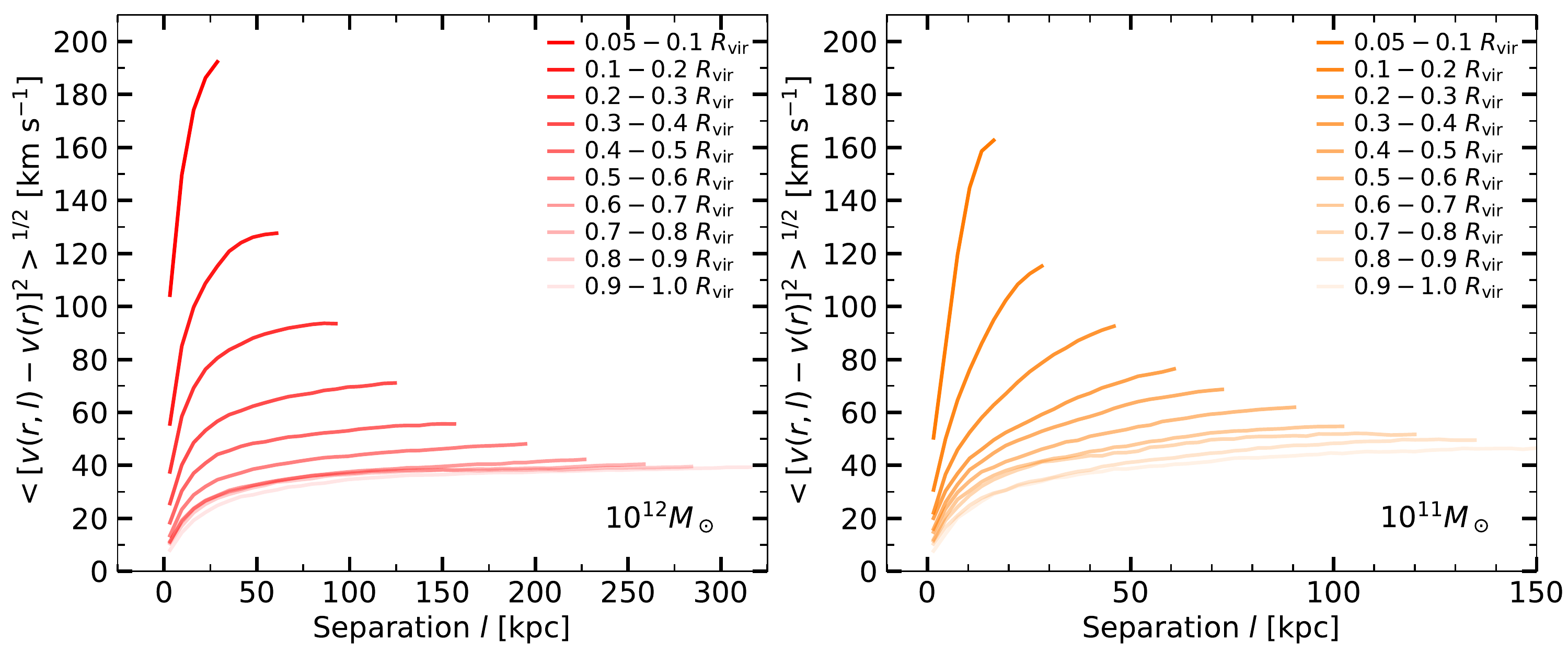}
\caption{The velocity structure function, defined as the average difference in velocity vectors between two gas parcels in the CGM, as a function of the separation between those parcels. Left panel shows the VSF for the $10^{12}M_\odot$ halo and right panel shows the VSF for the $10^{11}M_\odot$ halo. Only hot gas $T>10^{4.7}$ K is considered. Each curve shows the VSF for a different radial bin, where both gas parcels in each pair are located within the same radial bin, smaller radial bins are shown as darker curves and larger radial bins as lighter curves, as indicated in the legend. The VSF shows that velocity motions in both halos are correlated in ``patches" on scales of $\sim25$ kpc.}
\label{fig:vsf}
\end{minipage}
\end{figure*}

We measure the VSF as a function of the galactocentric radius. We calculate the VSF by randomly sampling 2000 pixels with $T>10^{4.7}$ K within each galactocentral radial bin (we wish to focus on only the hot gas), then calculating $[v(r,l)-v(r)]^2$ for every pair of pixels within each radial bin, where $v(r)$ is the velocity of a pixel in radial bin $r$ and $v(r,l)$ is the velocity of another pixel within the same radial bin $r$ at a separation $l$ from the first pixel in the pair. Note that the separation $l$ need not be in the radial direction. The VSF is then produced by averaging over every pair's $[v(r,l)-v(r)]^2$ within bins of $r$ and $l$ and taking the square root. We perform this calculation at every time snapshot and report the time-averaged values. The result is shown in Figure~\ref{fig:vsf}, where the VSF is plotted as a function of separation between pixels, and each curve shows the VSF for a given galactocentric radial bin. For ease of plot-reading, we do not show the $1\sigma$ time variation on each curve, which is roughly $30$ km s$^{-1}$ (20 km s$^{-1}$) in the smallest radial bin of the higher-mass (lower-mass) halo and decreases to 10 km s$^{-1}$ in the larger bins of both halos.

A lower value of the VSF implies more coherence in gas motions, which explains why all curves in Figure~\ref{fig:vsf} dip to low values at small separations $l$. At larger separations in the larger radial bins, the value of the VSF approaches a constant value that describes the overall motions of uncorrelated gas. This constant value is larger in the lower-mass halo, indicating overall more gas motions in the lower-mass halo, consistent with our findings above in Figure~\ref{fig:vr_radius}. In the smaller radial bins, the values of the VSF are larger, indicating that there is significantly more varied gas motions at smaller galactocentric radii, consistent with our findings that smaller radial bins have wider $v_\theta$, $v_\phi$, and $v_r$ distributions in Figure~\ref{fig:vel_hist}.

The shape of the VSF curves seems to indicate that there is, in fact, a characteristic scale of gas motions throughout the halo. There is a quick rise in the VSF from $l=0$ to $l\sim25$ kpc, followed by a slower rise to the constant value at larger $l$, in all radial bins other than the smallest bin, where the bin is too small for the VSF to reach the slowly-rising regime. This implies that the gas motions have an enhanced ``patchiness" at small scales, where the values of the VSF are smaller than the value of the uncorrelated motions at large separations. The characteristic size of the ``patches" can be read off of Figure~\ref{fig:vsf} as the separation at which the fast rise in VSF terminates, which is $l\sim25$ kpc in both halos. The power-law slope of the VSF within these ``patches" is $\sim0.3-0.4$, roughly consistent with that expected from Kolmogorov turbulence \citep{Kolmogorov1941}. The size of the ``patches" at $\sim25$ kpc indicates that this is the driving scale for the turbulence, or the maximum physical scale where the turbulent cascade to smaller physical scales begins. An interesting avenue for further study would be a more detailed examination of turbulence in the CGM using high resolution simulations.

\subsection{Pressure Support}
\label{sec:HSE}

The smooth behavior of the pressure distributions and the fact they do not vary significantly over time leads to the idea that these halos, especially the higher-mass halo, are close to hydrostatic equilibrium, where the thermal pressure supports the gas against gravity. To explore this further, we define a parameter that measures the degree of hydrostatic equilibrium,
\begin{align}
\alpha_\mathrm{HSE}&=-\frac{\nabla_r P_\mathrm{th}}{\rho \nabla_r \Phi} \label{eq:HSE} \\
&= -\frac{P_{\mathrm{th},i+1}-P_{\mathrm{th},i}}{\rho(\Phi_{i+1}-\Phi_i)} \nonumber
\end{align}
where $\Phi$ is the gravitational potential defined by the NFW dark matter halo, and $\nabla_r$ indicates the gradients of pressure and gravitational potential are taken in only the radial direction. The second equality shows how we calculate $\alpha_\mathrm{HSE}$ in practice, where the subscripts $i$ and $i+1$ indicate the average of the quantity in adjacent radial bins $i$ and $i+1$. If $\alpha_\mathrm{HSE}=1$, the gas is in hydrostatic equilibrium, if $\alpha_\mathrm{HSE}<1$, the gas is gravity-dominated and lacks the pressure support to hold it in place in the halo, so it may begin flowing inward, and if $\alpha_\mathrm{HSE}>1$, the gas has more than enough pressure to support itself against gravity and may be outflowing.

\begin{figure}
\centering
\includegraphics[width=\linewidth]{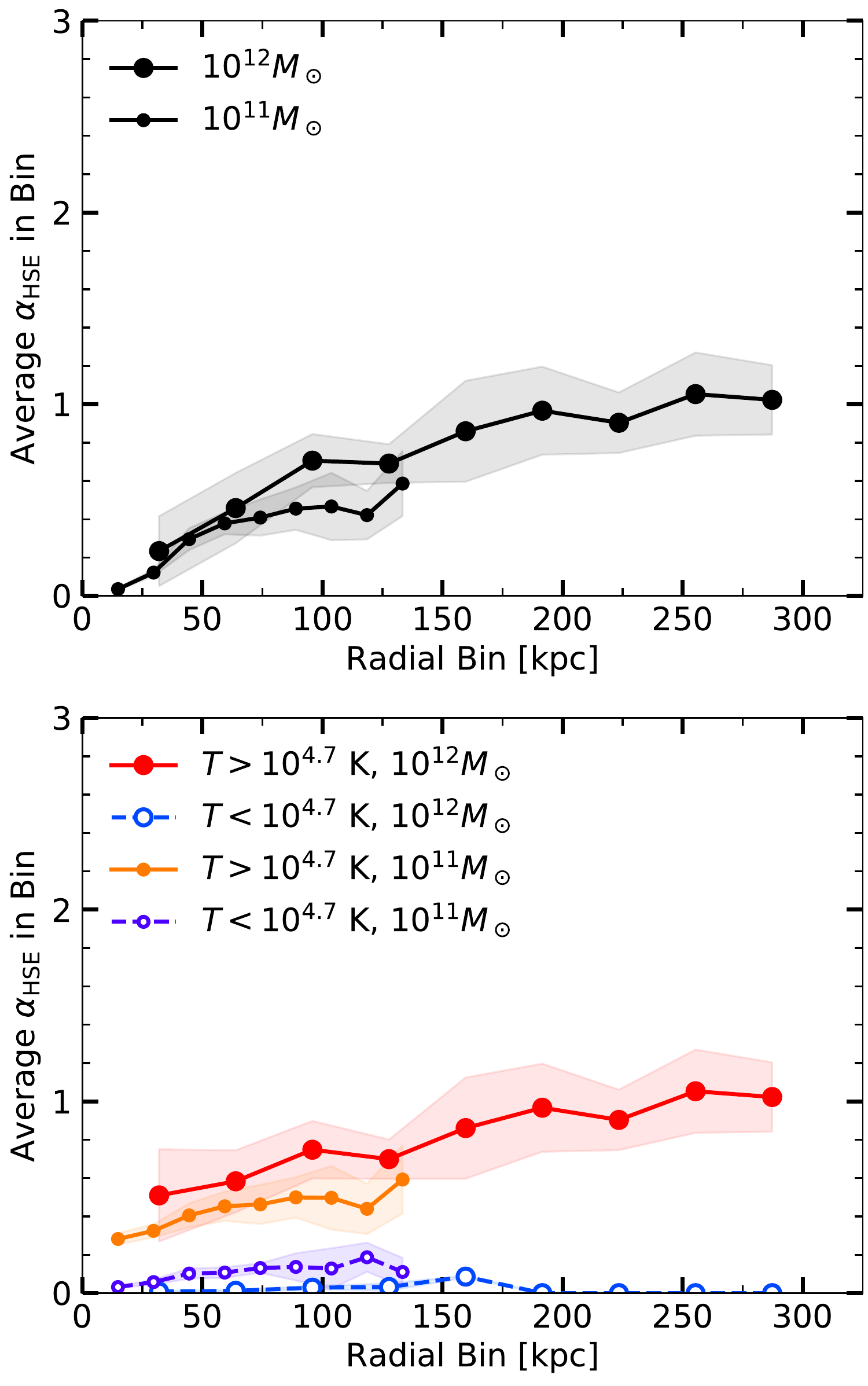}
\caption{The time-averaged $\alpha_\mathrm{HSE}$ (equation~\ref{eq:HSE}) as a function of radius for both the $10^{12}M_\odot$ (large circles) and $10^{11}M_\odot$ (small circles) halos. The top panel shows $\alpha_\mathrm{HSE}$ for all gas in each simulation, while the bottom panel shows $\alpha_\mathrm{HSE}$ when the gas is first separated by temperature at $T=10^{4.7}$ K. Shaded regions around curves indicate the one standard deviation time variance of the values. Note that the values in the top panel are the averages of the bottom panel only when weighted by the volume of gas in each temperature phase. Both halos do not have enough thermal pressure support to hold up their gas against gravity.}
\label{fig:HSE_thermal}
\end{figure}

Figure~\ref{fig:HSE_thermal} shows the time-averaged $\alpha_\mathrm{HSE}$ (only thermal pressure) as functions of radius, both for all gas and gas that has been split into hot ($T>10^{4.7}$ K) and cold ($T<10^{4.7}$ K). Both halos are closest to hydrostatic equilibrium in the outer radial bins, but under-supported in the inner regions of the halo close to the galaxy, despite these regions having the strongest thermal pressure (see Figure~\ref{fig:pres_radius}). The lower-mass halo is thermally under-supported throughout, but the higher-mass halo is in rough equilibrium for radii $\gtrsim 0.5R_\mathrm{vir}$. When the gas is split into low- and high-temperature regimes, then we see a striking difference in the thermal pressure support. The hot gas in both halos is closer to thermal pressure equilibrium ($\alpha_\mathrm{HSE}\approx1$) than the cold gas ($\alpha_\mathrm{HSE}\approx0.1$), although there is still an overall lack of pressure support.

To include the dynamic pressure support, in addition to the thermal pressure support, we define another parameter similar to $\alpha_\mathrm{HSE}$ (equation~\ref{eq:HSE}) but including dynamic pressure. Because contributions to the support against gravity from dynamic motions is now included, we drop the term ``hydrostatic" and instead call this new parameter just $\alpha_\mathrm{E}$ for ``equilibrium." $\alpha_\mathrm{E}$ is defined as:

\begin{align}
\alpha_\mathrm{E}&=-\frac{\nabla_r (P_\mathrm{th}+P_\mathrm{dyn})}{\rho \nabla_r \Phi - \nabla_r P_\mathrm{rad,in}} \label{eq:E} \\
&= -\frac{(P_{\mathrm{th},i+1} + P_{\mathrm{dyn},i+1}) - (P_{\mathrm{th},i} + P_{\mathrm{dyn},i})}{(\rho \Phi_{i+1} - P_{\mathrm{rad,in},i+1}) - (\rho \Phi_{i} - P_{\mathrm{rad,in},i})} \nonumber
\end{align}

where again, the second equality shows how we calculate $\alpha_\mathrm{E}$ in practice, where $i$ and $i+1$ denote adjacent radial bins. $P_\mathrm{dyn}$ can be the outward radial dynamic pressure, the tangential dynamic pressure, or the sum of these. In order to capture the contribution to pressure support of both the turbulent motions and the bulk flows, we consider $P_\mathrm{dyn}=P_\mathrm{rad,out}$ as a lower limit and $P_\mathrm{dyn}=P_\mathrm{rad,out}+P_\mathrm{tan}$ as an upper limit. $P_\mathrm{rad,out}$ is a lower limit because it includes only the radial direction of the isotropic turbulence, but $P_\mathrm{rad,out}+P_\mathrm{tan}$ is an upper limit because it then ``double counts" some of the turbulent pressure because $P_\mathrm{rad,out}$ contains a contribution from one of the three directions of the isotropic turbulence. Because we cannot separate the radial direction of the turbulence from the radial bulk flows, we are unable to exactly calculate the turbulent pressure support and bulk flow ram pressure support separately. In practice, the upper and lower limits are close to each other, so we are confident we can accurately constrain $\alpha_\mathrm{E}$. Note that the inward radial dynamic pressure is included in the denominator of equation~(\ref{eq:E}) because it acts in the same direction as gravity to oppose the supporting pressures of $P_\mathrm{th}$, $P_\mathrm{turb}$, and $P_\mathrm{rad,out}$. The negative sign before $P_\mathrm{rad,in}$ ensures it acts against the other pressures and in the same direction as gravity.

\begin{figure}
\centering
\includegraphics[width=\linewidth]{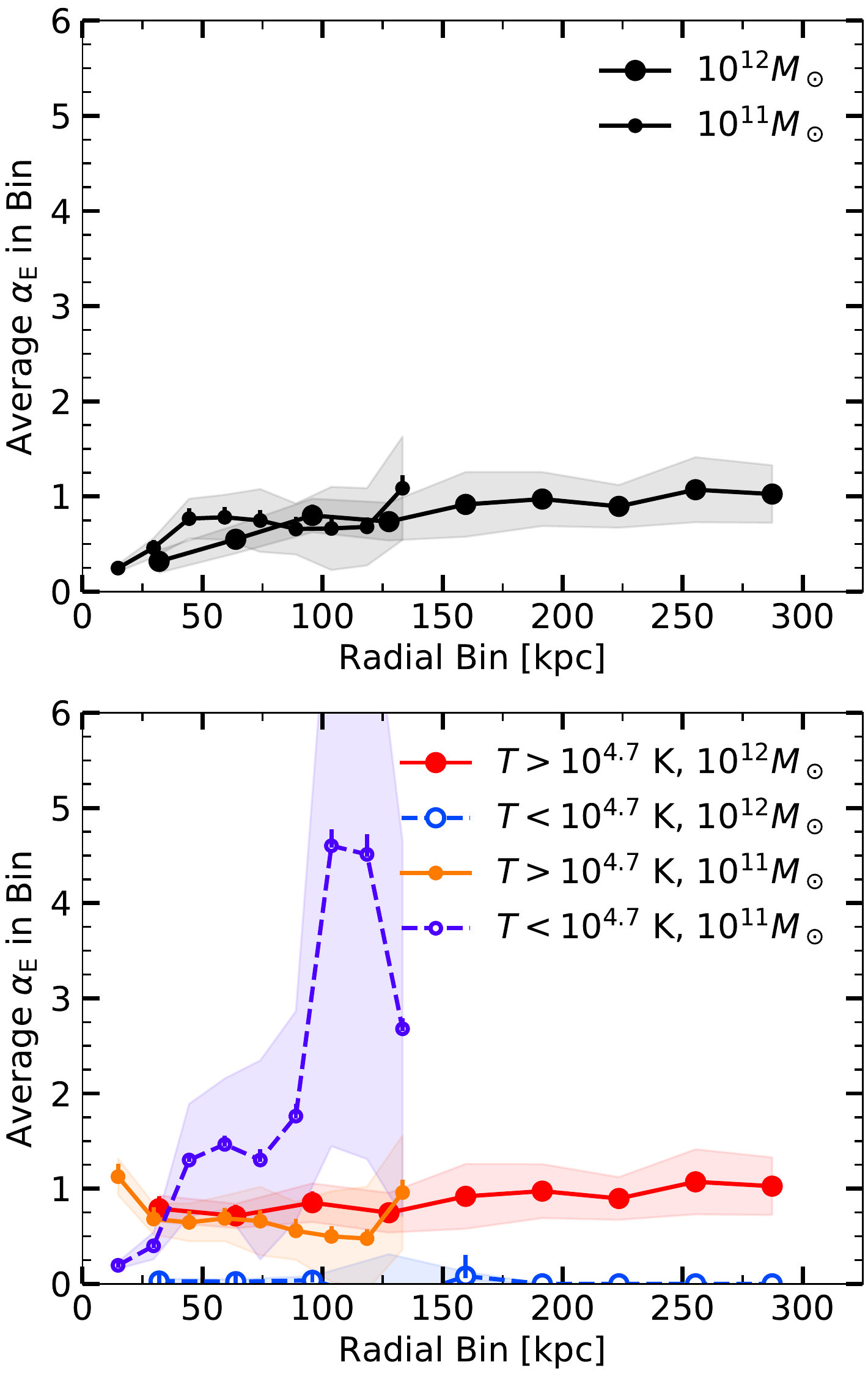}
\caption{The time-averaged $\alpha_\mathrm{E}$ (equation~\ref{eq:E}), now computed using the sum of thermal and dynamic pressures, as a function of radius for both the $10^{12}M_\odot$ (large circles) and $10^{11}M_\odot$ (small circles) halos. The connected points show the lower limit on $\alpha_\mathrm{E}$, where the dynamic pressure is calculated from only the radial direction of the dynamic pressure. Vertical lines extending upward from each point show the extent of the difference between this lower limit and the upper limit on $\alpha_\mathrm{E}$, which includes the tangential dynamic pressure as well (see text). The top panel shows $\alpha_\mathrm{E}$ for all gas in each simulation, while the bottom panel shows $\alpha_\mathrm{E}$ when the gas is first separated by temperature at $T=10^{4.7}$ K. Shaded regions around curves indicate the one standard deviation time variance of the values. Note that the values in the top panel are the averages of the bottom panel only when weighted by the volume of gas in each temperature phase. When dynamic pressure support is considered, the high-mass halo is roughly in pressure equilibrium while the low-mass halo is dominated by the pressure of outflows.}
\label{fig:E}
\end{figure}

Figure~\ref{fig:E} is similar to Figure~\ref{fig:HSE_thermal}, but now shows the upper and lower limits on $\alpha_\mathrm{E}$. With the inclusion of dynamic pressure support, the higher-mass halo is nearly in equilibrium ($\alpha_\mathrm{E}=1$) throughout, with only the innermost radial bins being under-supported. The lower-mass halo is still under-supported throughout, but is much closer to equilibrium than when only thermal pressure was considered and does not show as strong of a trend with radius (note the different vertical axis scales between Figures~\ref{fig:HSE_thermal} and~\ref{fig:E}).

\begin{figure}
\centering
\includegraphics[width=\linewidth]{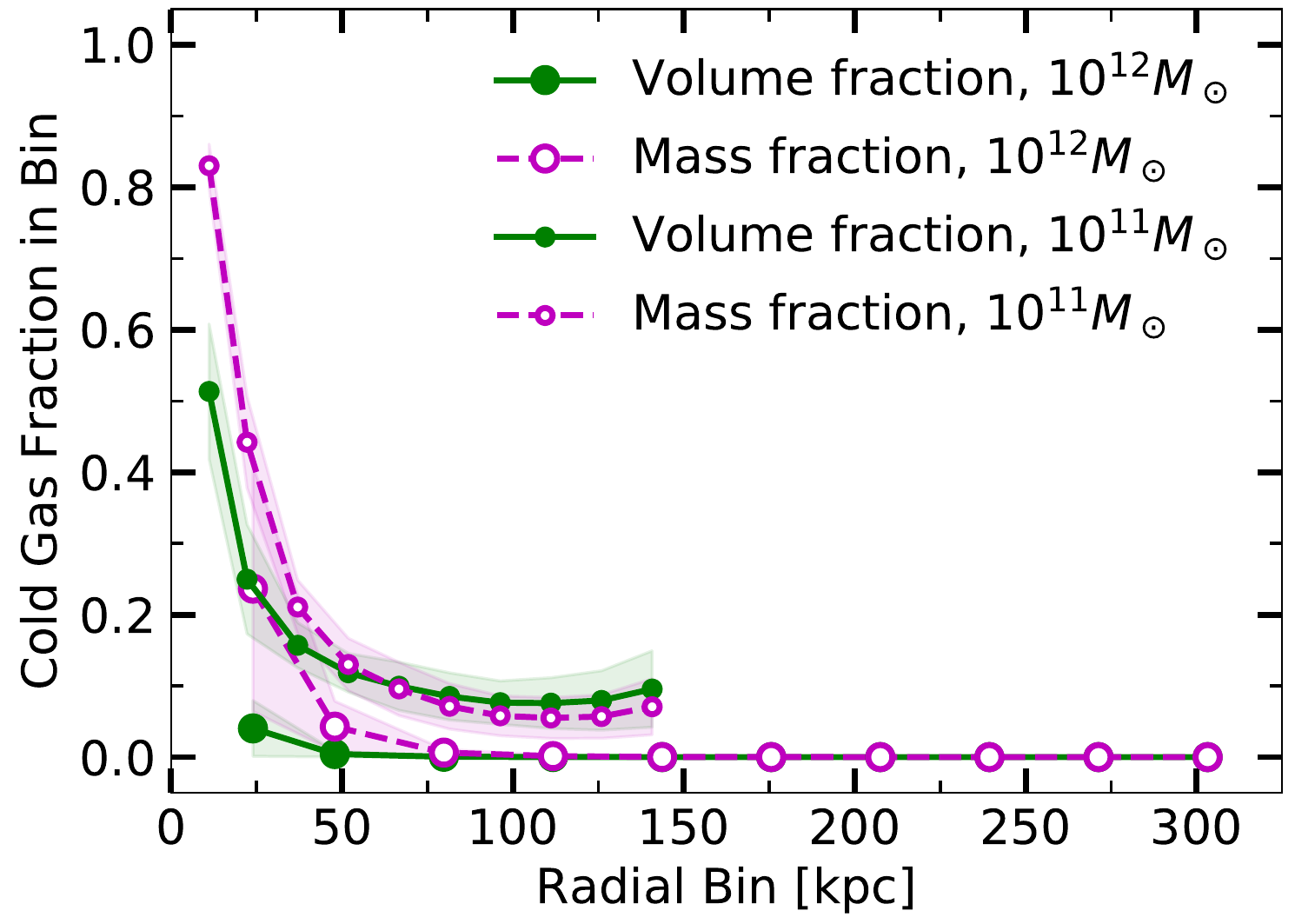}
\caption{The volume (green, filled points and solid lines) and mass (magenta, open points and dashed lines) fractions of cold gas within each radial bin as a function of radius. The volume and mass fractions in the higher-mass halo are plotted with large circles and are plotted with small circles in the lower-mass halo. Note that the hot and cold gas fractions sum to one. Both halos are dominated by hot gas in their outskirts, but cold gas becomes more prominent in the inner regions of the halo.}
\label{fig:m_v_frac}
\end{figure}

In order to better understand the picture presented by Figures~\ref{fig:HSE_thermal}-\ref{fig:E}, we plot the volume and mass fractions of the cold gas within each radial bin in Figure~\ref{fig:m_v_frac}. The hot gas dominates in both the higher- and lower-mass halos, and there is barely any cold gas at all past $\sim50$ kpc in the higher-mass halo. However, the lower-mass halo has substantially more cold gas, making up $\sim10\%$ of each radial bin by mass and volume past $\sim100$ kpc, increasing up to $\sim50\%$ or more in the smallest radial bins.

When the gas is split into hot and cold, the differences between the higher- and lower-mass halos become more stark. In the higher-mass halo, the hot gas is nearly perfectly in equilibrium throughout the halo, while the cold gas is under-supported throughout. In the lower-mass halo, neither the hot gas nor the cold gas are in equilibrium, although the hot gas is close to equilibrium but slightly under-supported throughout. The cold gas is under-supported in the inner parts of the lower-mass halo, passes through an equilibrium value at intermediate values, and then approaches more than enough support in the outskirts of the halo. The abundance of pressure support in the cool gas in the outskirts of the halo is provided by the outflow ram pressure, which dominates over other types of pressure in the cool gas (not shown, but see Figure~\ref{fig:pres_radius} for the relative strengths of the different pressures in all the gas). Combined with Figures~\ref{fig:vr_radius} and~\ref{fig:m_v_frac}, this suggests a picture for the lower-mass halo in which the majority of the hot gas is inflowing, but any pressurized warm outflows that are present rapidly cool as they expand away from the galaxy, contributing to the outflow ram pressure support of the cool gas on large scales.

In summary, the inclusion of dynamic pressure in calculating the full pressure support for the halo is crucial, especially for the pressure support of the cold gas that does not have much thermal pressure support and for determining the support of a low-mass halo. The pressure support of both hot and cold gas in the lower-mass halo comes primarily from its dynamic pressure, as seen by the drastic difference between $\alpha_\mathrm{HSE}$ and $\alpha_\mathrm{E}$ in the lower-mass halo. Only the higher-mass halo is in a steady state of pressure equilibrium at a majority of radii, which is provided nearly entirely by the hot gas (which also dominates the mass of the higher-mass halo). In the higher-mass halo, the cold gas is at high density, suggesting it is primarily located in small, dense clumps that have radiatively cooled out of the hot halo. In the lower-mass halo, there is much less difference between either the temperatures or the densities of the ``hot" and ``cold" gas, suggesting the $10^{4.7}$ K temperature split is somewhat arbitrary, and the warm gas is adiabatically and radiatively cooling en masse within outflows without forming cloudlets.

\section{Discussion}
\label{sec:discussion}

\subsection{Dependence of CGM Properties on Halo Mass}
\label{sec:halomass}

We have performed a detailed analysis of the properties of CGM gas within two simulations of galaxies hosted by $10^{12}M_\odot$ and $10^{11}M_\odot$ halos, and found a number of differences in the CGM between halos of these different masses. The lower-mass halo has a wider range of values within each radial bin in its thermal pressure distributions (Figure~\ref{fig:pres_radius}) and its degree of pressure equilibrium (both with and without dynamic pressure, Figures~\ref{fig:HSE_thermal} and~\ref{fig:E}). It is also more dominated by bulk flows than by turbulence (Figure~\ref{fig:time_fraction}) at small galactocentric radii. The CGM of halos of mass $\sim10^{11}M_\odot$ appears to be more dynamic than that of higher-mass halos. Combined with the fact that the gas in this halo is not in hydrostatic equilibrium, this paints a picture in which there is no steady-state of the CGM for low-mass halos.

On the other hand, the $10^{12}M_\odot$ halo follows the trends expected for a classical ``hot halo" in pressure equilibrium. The pressure distribution is well-behaved and there is only a small range of pressures within a given radial bin. The hot gas is roughly in thermal and dynamic pressure equilibrium at most large radii, while the cold gas is generally lacking thermal or dynamic pressure support.

Based on these results, the majority of the cold gas mass in the higher-mass halo is inflowing, while the hot gas is not participating in bulk flows and is supported by its thermal and turbulent pressure, a picture also expected for a hot gas halo from which cold clouds can condense to ``rain" onto the central galaxy \citep{Pizzolato2005,Soker2010,Gaspari2012,Voit2017}.

The lower-mass halo is reversed, with the majority of the cold gas mass outflowing and the hot gas mass inflowing, so perhaps the cold gas is sourced by the winds themselves as they adiabatically and radiatively cool \citep{Thompson2016}, which is supported by the finding that the outflow ram pressure is the dominant term in the pressure support of the cold gas in the outskirts of the low-mass halo.

The CGM of lower-mass halos should certainly not be considered in the same way as in higher-mass halos. It is dynamic and not in equilibrium, nor are the halo dynamics well-described by a model of cold gas condensing out of a hot medium. It is dominated by anisotropic bulk flows, not thermal gas properties, and is dependent on feedback parameters (see \S\ref{sec:feedback} below).

\subsection{Implications for Small-Box CGM Simulations}
\label{sec:smallbox}

Small-scale simulations of a portion of the CGM are necessary to achieve a spatial resolution approaching the cooling length which may be as small as $0.1$ pc \citep{Liang2018,McCourt2018}, and simulations of the full CGM, like those studied in this paper, are useful for setting the initial conditions for a small-box CGM simulation\footnote{Because the resolution in our simulations is not high enough to resolve the cooling length, we may be missing small-scale effects which impact the thermodynamics and/or dynamics of the gas. We do not think this affects our results because the thermal pressure of the cool gas is small compared to the hot gas, but small-box simulations would need to take this into account.}. The well-behaved and expected nature of the CGM of the $10^{12}M_\odot$ halo, i.e. a static hot halo in pressure equilibrium from which cold clouds condense from thermal instabilities, allows for easier modeling of a small patch of the CGM. Close to the central galaxy, there may be additional complications from strong galactic outflows, and especially the ``patchiness" of those outflows (Figure~\ref{fig:vsf}), but initializing a small-box simulation at a large galactocentric radius would require only information about the pressure support of the gas and its turbulent properties, provided the box is large enough, $\sim25$ kpc on a side, to capture the structure in the turbulent velocity ``patches." Such small-box simulations would be invaluable for exploring how thermal instability proceeds in the outer CGM of massive galaxies and the formation of cold gas.

The more anisotropic and non-equilibrium nature of the $10^{11}M_\odot$ halo's CGM creates a more complicated problem for small-box simulators. The CGM is dominated by bulk flows, both in and out, and there is not a strong difference between the hot and cold gas phases. There is significant dynamic pressure support and the isotropic turbulent velocities are the only well-behaved quantity. Both hot and cold gas can be both outflowing and inflowing, so initializing a small box to represent a patch of a lower-mass galaxy's CGM would require a careful choice of what stage of the CGM's evolution will be simulated. There are significant asymmetries that develop in the lower-mass halo, even in idealized simulations so a suite of small-box simulations at many locations within the halo would be necessary to understand the small-scale physical processes occurring everywhere within the CGM of a low-mass galaxy.

\subsection{Dependence on Feedback Parameters}
\label{sec:feedback}

In addition to the reference set of wind parameters for the $10^{11}M_\odot$ halo, we also examined three other combinations of wind speed $v_w$ and mass-loading $\eta$, with different energy efficiencies. The energy efficiency is defined as the ratio of $\eta v_w^2$ to $\dot M_\star c^2$ \cite[see][]{Fielding2017} and is $3\times10^{-6}$ for our reference simulation. The three other wind implementations are: one with the same mass-loading factor as our reference simulation ($\eta=5$) but with a lower $v_w^2=v_\mathrm{esc}^2$ (instead of the $v_w^2=3v_\mathrm{esc}^2$ of our reference simulation), which has an energy efficiency of $1\times10^{-6}$, and two with a lower mass-loading factor of $\eta=0.3$, with $v_w^2=4.5v_\mathrm{esc}^2$ (energy efficiency $0.3\times10^{-6}$) or $v_w^2=9v_\mathrm{esc}^2$ (energy efficiency $0.6\times10^{-6}$). All other initial conditions were kept the same, and we calculated distributions within radial bins of the thermal pressures, radial velocities, and $\alpha_\mathrm{HSE}$ and $\alpha_\mathrm{E}$ (equations~\ref{eq:HSE} and~\ref{eq:E}), and calculated the time-averaged median of these distributions, as before. We also calculated the time-averaged mean tangential and radial dynamic pressures within the radial bins.

\begin{figure*}
\begin{minipage}{175mm}
\centering
\includegraphics[width=\linewidth]{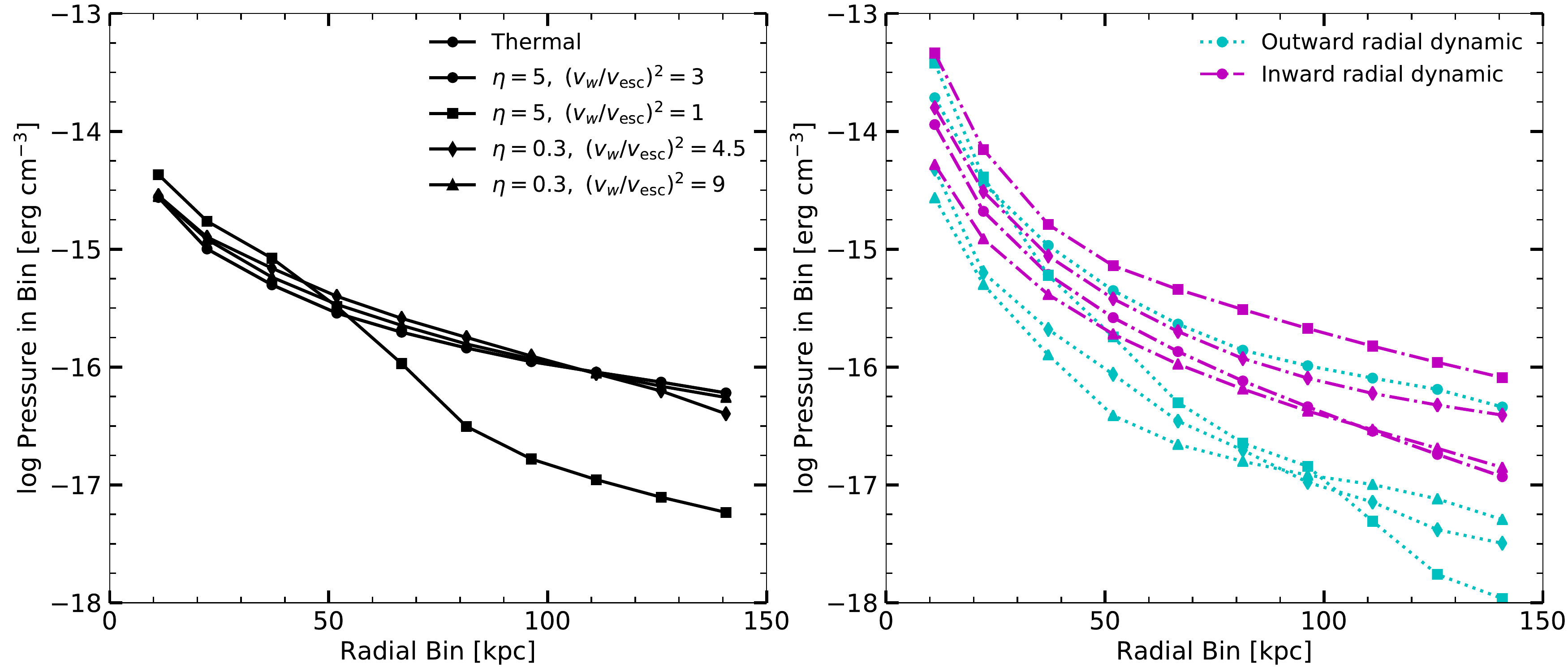}
\caption{The time-averaged median thermal pressure (black symbols and solid lines, left panel) and mean radial dynamic pressure (outward: cyan symbols and dotted lines, inward: magenta symbols and dash-dot lines, right panel) as functions of radius for $10^{11}M_\odot$ halos with different sets of wind parameters: the reference case with $\eta=5$ and $v_w^2=3v_\mathrm{esc}^2$ (circles), a case with $\eta=5$ and $v_w^2=v_\mathrm{esc}^2$ (squares), a case with $\eta=0.3$ and $v_w^2=4.5v_\mathrm{esc}^2$ (diamonds), and a case with $\eta=0.3$ and $v_w^2=9v_\mathrm{esc}^2$ (triangles). Low-mass halos implemented with different feedback parameters maintain similar thermal pressure profiles, but have different dynamic pressure profiles.}
\label{fig:pres_radius_windtypes}
\end{minipage}
\end{figure*}

Figure~\ref{fig:pres_radius_windtypes} shows the median of the thermal pressure and mean of the outward  and inward radial dynamic pressures for all four $10^{11}M_\odot$ simulations, split into two panels for ease of plot-reading. We do not plot the tangential dynamic pressures, as these are sub-dominant pressures in this halo, as we saw in Figure~\ref{fig:pres_radius}. As before, we see the radial dynamic pressures are typically larger than the thermal pressures in the inner regions of the halo, while the two types of pressure are more similar in the outskirts of the halo.

There is not a large difference in the medians of the thermal pressures (black points with solid lines) between different sets of wind parameters, with the exception of the one simulation with $\eta=5$ and $v_w^2=v_\mathrm{esc}^2$ (squares), which has the lowest wind speed out of the sets of wind parameters explored here. The wind does not travel very far into the halo and the larger radial bins for this simulation do not contain any wind material at all. The gas that is accreting onto the halo at these large radii is cold and thermally under-supported. At small radii where the outflows are present, there is no difference in the thermal pressure median between this set of wind parameters and any of the others.

The only simulation in which the outward radial dynamic pressure is equal to or greater than the inward radial dynamic pressure is the reference simulation. This is the only simulation to exhibit large-scale outflows. The other wind implementations produce inflow-dominated halos where the inward radial dynamic pressure is significantly stronger than the outward radial dynamic pressure, especially in the outskirts of the halo.

\begin{figure}
\centering
\includegraphics[width=\linewidth]{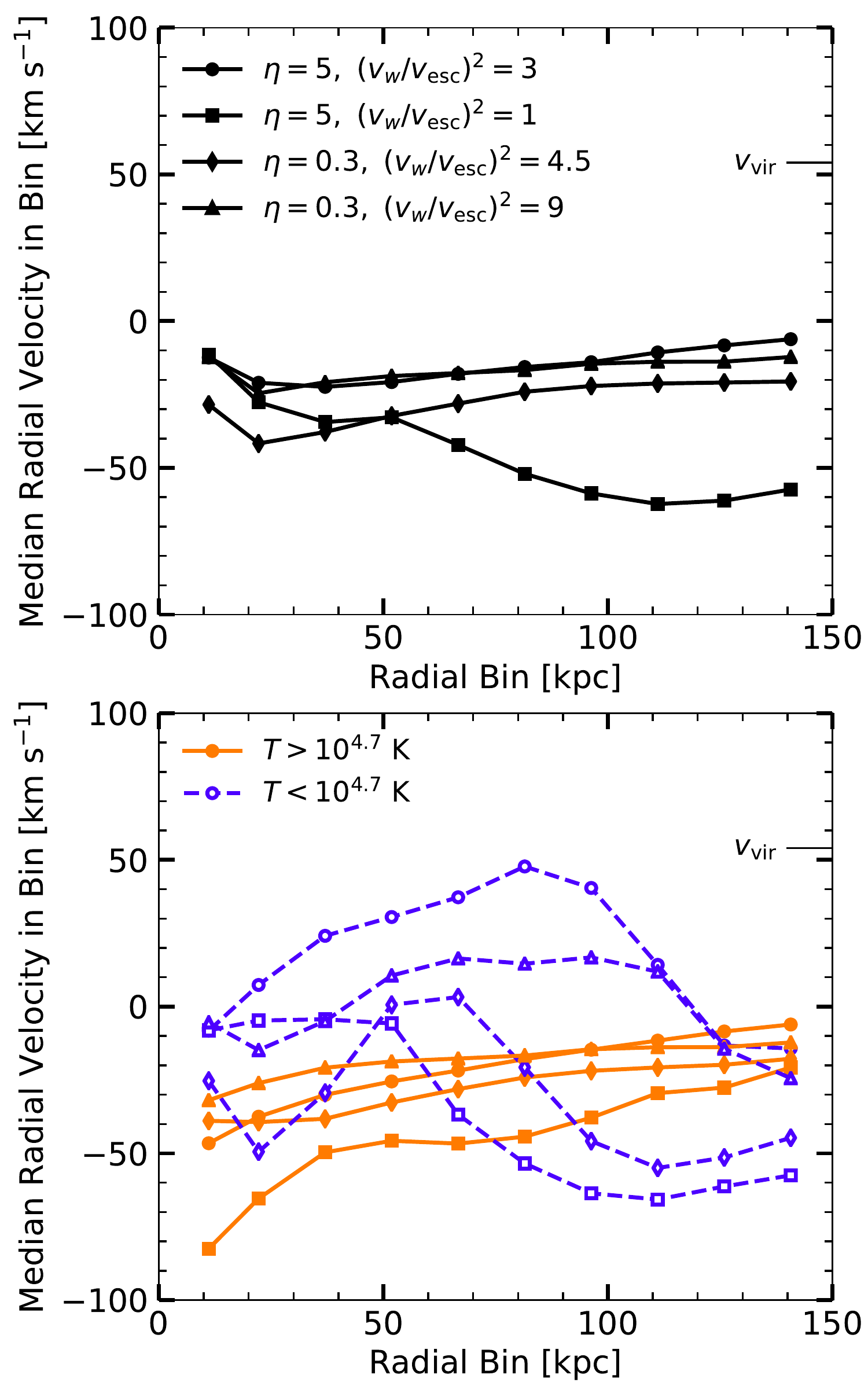}
\caption{The time-averaged median radial velocity as a function of radius for all $10^{11}M_\odot$ halos with different sets of wind parameters: the reference case with $\eta=5$ and $v_w^2=3v_\mathrm{esc}^2$ (circles), a case with $\eta=5$ and $v_w^2=v_\mathrm{esc}^2$ (squares), a case with $\eta=0.3$ and $v_w^2=4.5v_\mathrm{esc}^2$ (diamonds), and a case with $\eta=0.3$ and $v_w^2=9v_\mathrm{esc}^2$ (triangles). The top panel shows radial velocities for all gas in each simulation, while the bottom panel shows the median of the radial velocity distributions when the gas is first separated by temperature at $T=10^{4.7}$ K. Note that the values in the top panel are the averages of the bottom panel only when weighted by the mass of gas in each temperature phase. The virial velocity of this halo is marked on the right axis in each panel for comparison. Different feedback parameters affect the overall motions of the gas in the low-mass halo, but all sets of parameters lead to an inflow-dominated halo.}
\label{fig:vr_radius_windtypes}
\end{figure}

Figure~\ref{fig:vr_radius_windtypes} shows the median of the radial velocity in these four simulations for all gas (top panel) and gas split into hot and cold (bottom panel) at $10^{4.7}$ K. All sets of wind parameters produce an inflow-dominated halo, similar to the reference case. The simulation with $\eta=5$ and $v_w^2=v_\mathrm{esc}^2$ (squares) has faster inflows at large radius than the other simulations for the same reason as it is under-supported: the outflows do not travel to large radii so there is only the cosmological accretion at these large radii. Similarly, the simulation with $\eta=0.3$ and $v_w^2=4.5v_\mathrm{esc}^2$ (diamonds) is slightly more inflow-dominated than the others, again because the outflows in this simulation do not travel all the way to the edge of the simulation domain, and end up returning to the central galaxy.

When the gas is split by temperature, the general trend of the hot gas smoothly inflowing while the cold gas has more variation in radial velocity is observed for all simulations as in the reference case. The reference simulation (circles) is the only to exhibit outflows on average. The others have highly anisotropic, cold outflows and inflows as evidenced by non-monotonic trends in $v_r$ over the range of radii.

\begin{figure}
\centering
\includegraphics[width=\linewidth]{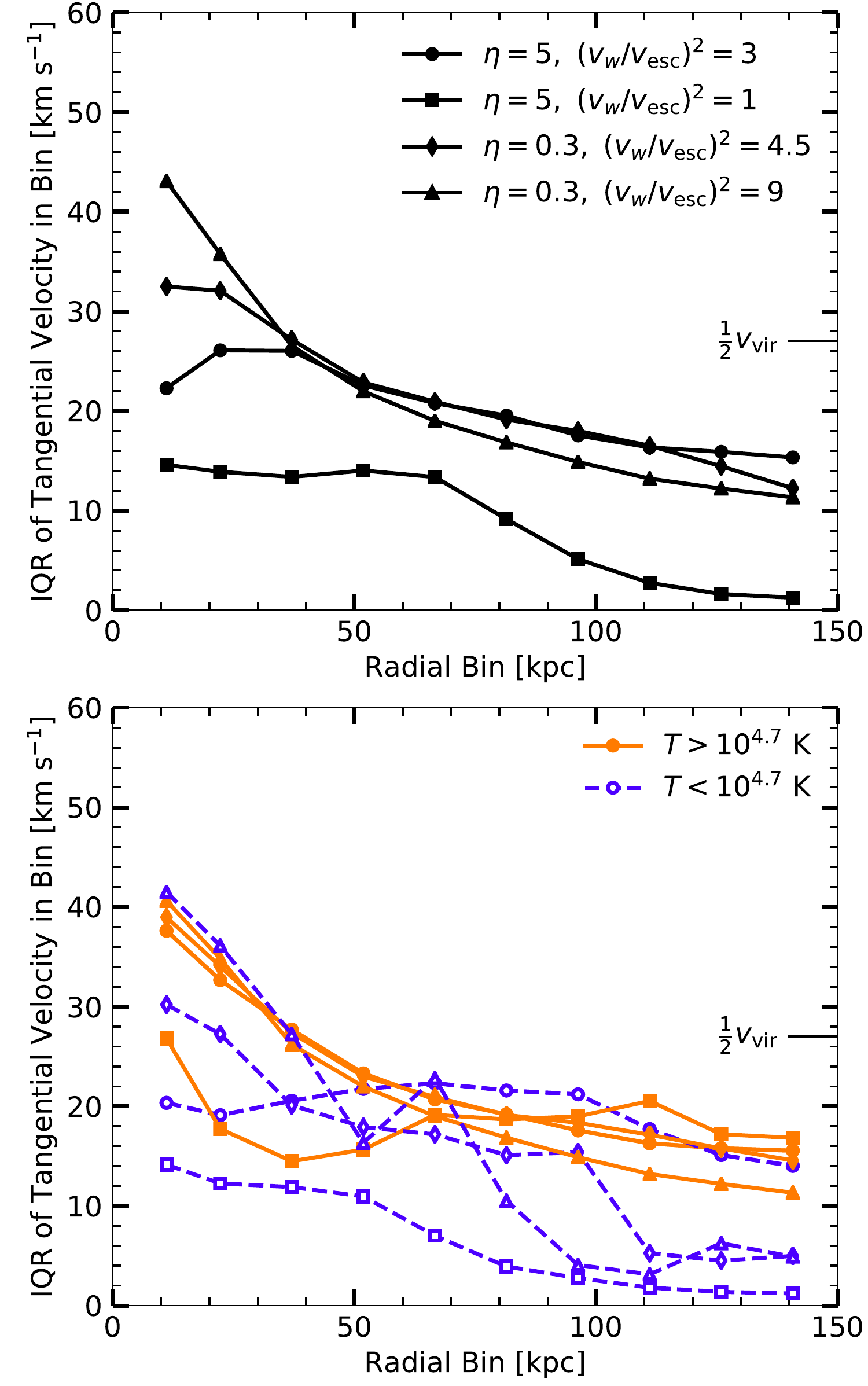}
\caption{The IQR of the tangential velocity distribution as a function of radius for all $10^{11}M_\odot$ halos with different sets of wind parameters: the reference case with $\eta=5$ and $v_w^2=3v_\mathrm{esc}^2$ (circles), a case with $\eta=5$ and $v_w^2=v_\mathrm{esc}^2$ (squares), a case with $\eta=0.3$ and $v_w^2=4.5v_\mathrm{esc}^2$ (diamonds), and a case with $\eta=0.3$ and $v_w^2=9v_\mathrm{esc}^2$ (triangles). Top panel shows the tangential velocity IQRs for all gas and the bottom panel shows the IQR when the gas is first separated by temperature at $T=10^{4.7}$ K. Note that the values in the top panel are the averages of the bottom panels only when weighted by the mass of gas in each temperature phase. Half of the virial velocity value for this halo is marked on the right axes for comparison. Stronger feedback implementations drive faster turbulence in the low-mass halo.}
\label{fig:vt_radius_windtypes}
\end{figure}

Figure~\ref{fig:vt_radius_windtypes} shows the IQR of tangential velocities, which traces turbulent velocities, in the simulations with different wind implementations, both for all gas (top panel) and temperature-split gas (bottom panel). Most wind implementations produce similar turbulent velocities to the reference simulation, but the $\eta=5$, $(v_w/v_\mathrm{esc})^2=1$ simulation (squares) is again an outlier with very low turbulence. Because the galactic outflows in this simulation are massive and slow, their impact on the halo is limited and they do not trigger the same degree of turbulent motions that the other wind implementations do. When the gas is split into hot and cold (bottom panel), we see both the hot and cold gas in this simulation have smaller tangential velocity IQRs than the other wind implementations. Like in the reference simulation, the other wind implementations have faster turbulent motions in the warm gas than the cool gas at small and large radii, but similar turbulent velocities at intermediate radii.

\begin{figure}
\centering
\includegraphics[width=\linewidth]{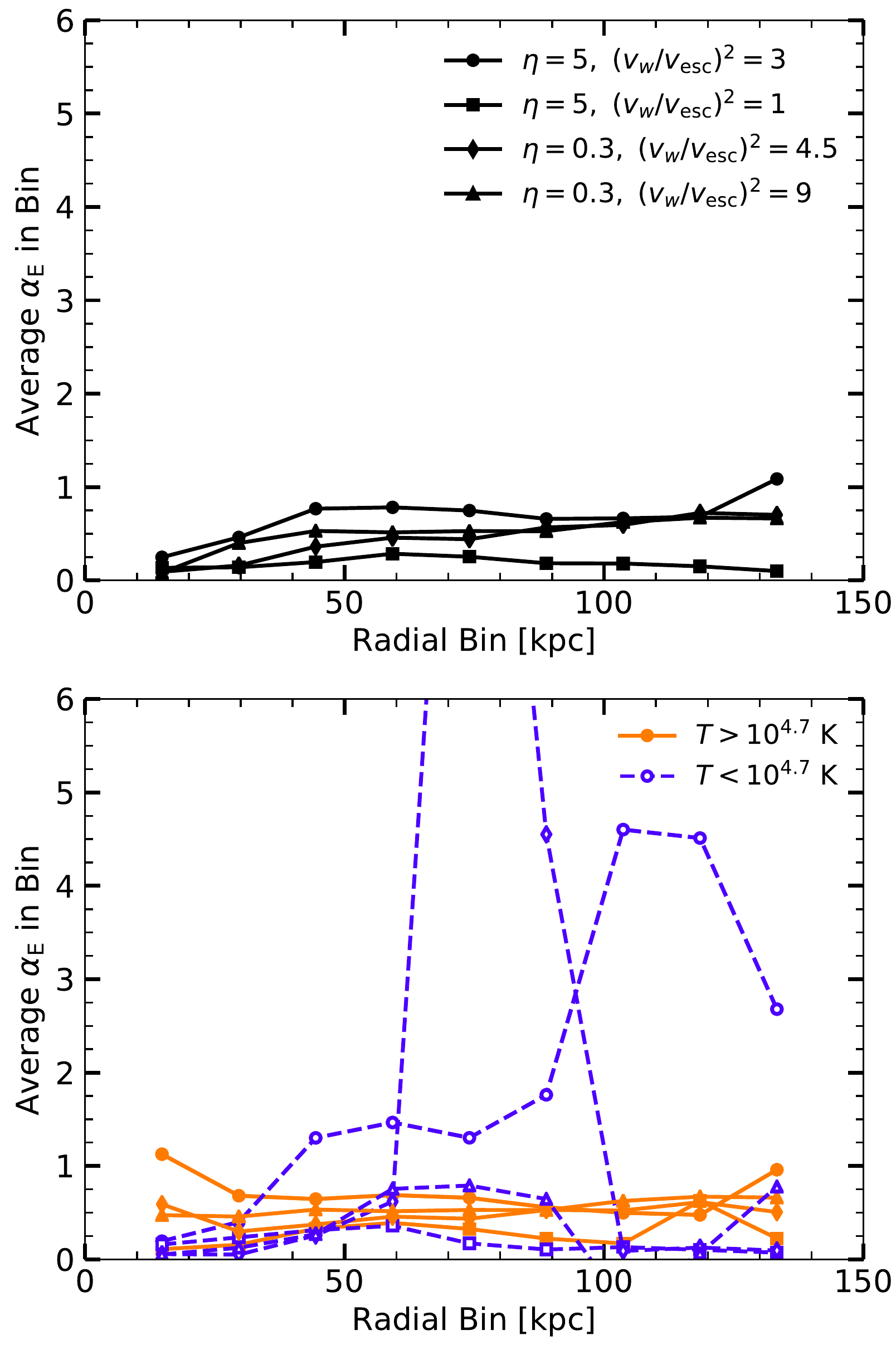}
\caption{The time-averaged $\alpha_\mathrm{E}$ (equation~\ref{eq:E}), computed using the sum of thermal and radial dynamic pressures, as a function of radius for $10^{11}M_\odot$ halos with different sets of wind parameters: the reference case with $\eta=5$ and $v_w^2=3v_\mathrm{esc}^2$ (circles), a case with $\eta=5$ and $v_w^2=v_\mathrm{esc}^2$ (squares), a case with $\eta=0.3$ and $v_w^2=4.5v_\mathrm{esc}^2$ (diamonds), and a case with $\eta=0.3$ and $v_w^2=9v_\mathrm{esc}^2$ (triangles). The top panel shows $\alpha_\mathrm{E}$ for all gas in each simulation, while the bottom panel shows $\alpha_\mathrm{E}$ when the gas is first separated by temperature at $T=10^{4.7}$ K. Note that the values in the top panel are the averages of the bottom panel only when weighted by the volume of gas in each temperature phase. Regardless of feedback parameters, the low-mass halo is not in pressure equilibrium.}
\label{fig:E_radius_windtypes}
\end{figure}

Finally, we consider the degree of pressure equilibrium in the halos with these different wind parameters. Figure~\ref{fig:E_radius_windtypes} shows the time-average of the lower limit on $\alpha_\mathrm{E}$ (equation~\ref{eq:E}) calculated using the sum of thermal and radial dynamic pressures for the gas in all $10^{11}M_\odot$ simulations, for all gas (top panel) and gas that is split in temperature into hot and cold (bottom panel). We focus on the radial dynamic pressure for these simulations, as it is dominant over the tangential dynamic pressure. We do not plot vertical lines from each point showing the difference in $\alpha_\mathrm{E}$ between using $P_\mathrm{th}+P_\mathrm{rad,out}$ and $P_\mathrm{th}+P_\mathrm{rad,out}+P_\mathrm{tan}$ as we did in Figure~\ref{fig:E} for ease of plot-reading. The difference in $\alpha_\mathrm{E}$ calculated these two ways is very small in these cases.

All wind implementations produce halos that are under-supported in the inner regions of the halo is considered, but some of the wind implementations produce strong enough bulk outflows that the halo starts to approach pressure equilibrium in the outskirts of the halo, like the reference simulation. There is a general trend of under-supported hot gas at all radii, but the over-pressurized cold gas at large radii that was found in the reference simulation is only found with one other set of wind parameters. Again, we see the simulation with the lowest wind speed (squares) produces the most under-supported and inflow-dominated halo. The simulations with the highest wind speeds (triangles and diamonds) produce halos closest to the pressure support scenario seen in the reference simulation, but do not show the overly-supported cool gas at nearly all radii present in the reference. This implies that winds must be both sufficiently mass-loaded and sufficiently fast to produce pressure equilibrium in a low mass halo, and not even the reference simulation with the strongest winds can fully reach equilibrium. Additional feedback parameters not examined here, such as a feedback scenario with a much higher energy efficiency, might lead to a different picture of pressure support and gas motions in the lower-mass halo and should be explored further.

Overall, we find that changing the wind parameters has the largest effect on the dynamics of the gas in the CGM. We reiterate the conclusion from \S\ref{sec:smallbox} that the lower-mass halo is not only more dynamic and further from equilibrium than the higher-mass halo, but its dynamics and support are also strongly dependent on the properties of the wind, which only serves to make initializing small-box simulations of the CGM in low-mass halos more difficult.

\section{Summary and Conclusions}
\label{sec:summary}

We have characterized the circumgalactic medium of two simulated galaxy halos of dark matter mass $10^{12}M_\odot$ and $10^{11}M_\odot$. These simulations are of idealized, isolated galaxies with spherically symmetric potentials and cosmological accretion, and do not model the galaxy itself nor any cosmological evolution, but are high-resolution and include radiative cooling in photoionization equilibrium and galactic winds parameterized by a mass loading factor and a wind velocity.

Our main findings are as follows:
\begin{enumerate}
\item The thermal pressure in the CGM is well-described by roughly log-normal distributions that can be characterized by a median and interquartile range. Higher-mass halos have narrower distributions of thermal pressure than lower-mass halos.
\item Distributions of radial and tangential velocities of all gas within radial bins show that velocities tangential to the radial direction show properties consistent with turbulence, while the radial velocity distribution contains information on inflows, outflows, and turbulence. The velocities of both outflows and turbulence are larger in higher-mass halos; outflows are faster by design and drive stronger turbulence. The radial and tangential velocities in both halos are proportional to the virial velocity of each halo.
\item Both high- and low-mass halos contain predominantly inflowing gas when all gas is considered, but when the gas is split into high temperature ($T>10^{4.7}$ K) and low temperature ($T<10^{4.7}$ K), cold gas in high (low) mass halos is inflowing (outflowing), and hot gas in high (low) mass halos is static (inflowing).
\item The velocity structure function of the hot gas reveals that both halos' gas motions are patchy with structures of size $\sim25$ kpc. Within these patches, the slope of the velocity structure function is consistent with that expected from Kolmogorov turbulence.
\item When only thermal pressure is considered, both halos are not fully supported against gravity, but the higher-mass halo approaches hydrostatic equilibrium in the outskirts of the halo, especially in its hot gas. The hot gas provides nearly all of the thermal pressure support while the cold gas is extremely under-supported.
\item When radial (ram) and tangential (turbulent) dynamic pressures are included in the calculation of pressure support, the higher-mass halo is nearly in perfect equilibrium, provided by its hot gas, while the cold gas is under-supported. The lower-mass halo is still not in equilibrium, and instead the warm gas is close to equilibrium, but still under-supported, at all radii, while the cold gas is under-supported at small radii and pressurized at large radii. This supports a picture of low-mass warm outflows adiabatically and radiatively cooling as they expand to large radii in the lower-mass halo.
\item In analyzing simulations with different feedback parameters, we found variations in the mass loading and wind velocity produce the largest effects on the dynamics in the low-mass halo. Galactic winds with low wind speeds or low mass loading factors tend to produce inflow-dominated halos with little pressure support while those with both higher mass loading and wind speed are closer to pressure equilibrium.
\end{enumerate}

The high-mass halo's CGM follows the expected picture of a hot gas halo in thermal pressure equilibrium out of which cold gas condenses and rains onto the galaxy, but the low-mass halo does not have a static hot gas halo, is not in pressure equilibrium, and is not consistent with condensing cold gas. In addition, the properties of the low-mass halo are dependent on the galactic wind parameters. These findings imply that simulators who wish to perform small-box simulations of a patch of the CGM in order to accurately trace thermal instability and condensation at high resolution must make careful decisions about how to initalize their simulations, as the properties of any given patch of the CGM in a low-mass halo vary significantly with location in the halo and galactic wind properties, and the CGM has significant ``patchiness" in its velocity structure. We have shown that there is no such thing as an ``average" CGM in low-mass halos.

The idealized nature of these simulations is useful for understanding the physical processes that drive the evolution of the CGM, but there are many areas for improvement. Cold cosmological accretion is typically not perfectly spherically-symmetric, but may pierce through the CGM in filaments. The typical galaxy goes through many mergers during its evolution that may drastically change its CGM. Even in such ideal cases as studied here, the multiphase nature of the CGM requires careful analysis of not just the average properties of all gas, but the properties of each individual phase, a statement which is likely to only become more true when more complicating physics are included in the simulation. This paper provides a component of a physically-motivated characterization of the complete nature of the CGM and a method for analyzing other, less idealized simulations that may produce different results. In particular, the FOGGIE simulations of \citet{Peeples2019} are also Eulerian, allowing higher resolution in the diffuse CGM than particle-based simulations, and are fully cosmological. A similar analysis as presented here is underway on the FOGGIE simulations and will provide a deeper understanding of the dynamics and pressure support of the CGM in a fully cosmological context.

\section*{Acknowledgments}

The authors thank the referee for useful suggestions that improved the quality of the paper. CL thanks Todd A. Thompson and Adam K. Leroy for helpful discussions. CL is supported in part by NSF Grant 1516967. GLB acknowledges support from NSF grants AST-1615955 and OAC-1835509 and NASA grant NNX15AB19G. This work was initiated as a project for the Kavli Summer Program in Astrophysics held at the Center for Computational Astrophysics of the Flatiron Institute in 2018. The program was co-funded by the Kavli Foundation and the Simons Foundation. We thank them for their generous support. Some computations in this paper were carried out on the NASA Pleiades supercomputer and the NSF supported XSEDE program.

\end{document}